\documentclass[
twocolumn,
superscriptaddress,
bibnotes,footinbib,aps,
prb, longbibliography,
floatfix]{revtex4-2}
\usepackage[normalem]{ulem}
\usepackage{amsmath}
\usepackage{natbib}
\usepackage{hyperref}
\usepackage{amscd,latexsym}
\usepackage{mathrsfs}
\usepackage{graphicx}
\usepackage{epstopdf}
\usepackage{amsfonts}
\usepackage{exscale}
\usepackage{dcolumn}
\usepackage{bm}
\usepackage{color}

\begin{document}


\title{Giant hysteretic magnetoresistance accompanying the Mott transition and spin-glass state in organic metal}

\author{P.\,D.~Grigoriev}
\affiliation{L.\,D.~Landau Institute for Theoretical Physics, 142432, Chernogolovka, Russia}
\affiliation{
National University of Science and Technology ''MISIS'', 119049, Moscow, Russia} 
\affiliation{National Research University Higher School of Economics, Moscow 101000, Russia}
\author{S.\,I. Pesotskii}
\affiliation{FRC of Problems of Chemical Physics and Medicinal Chemistry RAS, Chernogolovka, Moscow region, 142432, Russia}
\author{R.\,B. Lyubovskii}
\affiliation{FRC of Problems of Chemical Physics and Medicinal Chemistry RAS, Chernogolovka, Moscow region, 142432, Russia}
\author{S.\,A. Torunova}
\affiliation{FRC of Problems of Chemical Physics and Medicinal Chemistry RAS, Chernogolovka, Moscow region, 142432, Russia}
\author{D.\,S.~Lyubshin}
\affiliation{L.\,D.~Landau Institute for Theoretical Physics, 142432, Chernogolovka, Russia}
\author{V.\,N.~Zverev}
\affiliation{Institute of Solid State Physics RAS, Chernogolovka, Moscow region, 142432, Russia}

\begin{abstract}
The giant magnetoresistance with a huge hysteresis is observed in the organic metal $\kappa$-(BEDT-TTF)$_2$Hg(SCN)$_2$Br at low temperature in a pressure interval around 3$\,$kbar of a width $\sim 1$\,kbar. The hysteretic magnetoresistance is isotropic with respect to the direction of magnetic field, which excludes the orbital effect of magnetic field as its origin. The observed temperature and magnetic-field dependence of this hysteresis and of its relaxation time indicates the strong influence of spin-glass state on magnetoresistance. Although a quantitative theory of this effect, originating from strong electronic correlations, requires complex numerical calculations, we suggest its explanation and a simple model which qualitatively describes the observed magnetoresistance behavior and shows a strong charge-spin entanglement. The proposed effect suggests a new class of extreme magnetoresistance mechanisms. 
\end{abstract}

\date{\today}

\maketitle

\section{Introduction}

The giant magnetoresistance (GMR) is a subject of huge research activity because of interesting physics and promising applications for magnetic field sensors (e.g., see \cite{NobelLectureGMR2008,Ennen2016} for reviews). There are many possible mechanisms of GMR. The magnetic multilayer structures, where magnetization is controlled by an external field, give GMR which is already applied in technology \cite{NobelLectureGMR2008,Ennen2016}. Another GMR mechanism associated with a ferromagnetic-to-paramagnetic phase transition, known as colossal magnetoresistance (CMR) \cite{Ramirez1997}, is actively investigated for already thirty years \cite{Ramirez1997,Tafra2025,Balguri2025}. The extremely large magnetoresistance (XMR) includes several different mechanisms \cite{Niu2022}, most pronounced at low temperatures. Usually, all these mechanisms can be understood in the one-electron approximation, i.e. when the electron-electron (e-e) interaction can be neglected.

Another important area of the condensed matter physics deals with strong electronic correlations. The high-Tc superconductivity, charge- and spin-density waves, Mott metal-insulator transitions and many other interesting and promising phenomena cannot be properly described without the e-e interaction. Usually, the GMR and strongly-correlated electronic systems are not directly coupled. Below we propose and investigate the system where GMR originates from strong electronic correlations. We report both the detailed experimental analysis and a theoretical description of GMR originating from the Mott transition. 

The organic metal $\kappa$-(BEDT-TTF)$_2$Hg(SCN)$_2$Br belongs to the well-known family of quasi-two-dimensional conductors, which are single crystal samples of cation - radical salts in which flat organic molecules of BEDT-TTF form cationic layers alternating with inorganic anionic ones. When the salt is formed, an electron from every second BEDT-TTF molecule goes to the anion. The remaining electrons, moving between BEDT-TTF molecules, form a three-quarters filled hole metal band inside the layers, the conductivity of which is several orders of magnitude higher than the conductivity between the layers \cite{Ishiguro1998}. In samples with packaging of BEDT-TTF molecules in the $\kappa$-type layer, such molecules form dimers that create a geometrically frustrated flat triangular lattice, which leads to the splitting of the conduction band into a completely filled and half-filled one \cite{2}.  This circumstance, as well as the narrow conduction band, $W\sim 0.2\,$eV \cite{Pesotskii2022}, create the necessary conditions $U/W >$ 1 for the transition of the metal $\kappa$-(BEDT-TTF)$_2$Hg(SCN)$_2$Br to the Mott-insulator state, where $U$ is the Coulomb repulsion energy on one dimmer. Indeed, in this metal, during cooling, a metal-insulator transition (MIT) was detected in the temperature dependence of the resistance at $T_{MI}\sim 90\,$K \cite{3}, and the state below the transition temperature was interpreted as a state similar to a Mott insulator. The low temperature state was studied by various methods at ambient pressure \cite{3,Hemmida2018,5,6}. Evidence was found of a spin glass state in the spin system at temperatures below 20$\,$K \cite{Hemmida2018}, and a dipole liquid state in the charge system \cite{Hassan2018,Hemmida2018}. Both of these states are a consequence of the original lattice frustration. At $T\sim$ 90$\,$K the metal $\kappa$-(BEDT-TTF)$_2$Hg(SCN)$_2$Br also undergoes a structural phase transition \cite{Pesotskii2022} from a monoclinic syngony above 90$\,$K to a triclinic one below 90$\,$K. This transition is accompanied by a pronounced hysteresis in the temperature dependence of the resistance. Theoretical calculations of the band structure showed \cite{Pesotskii2022} that the triclinic lattice corresponds to a new but also metallic electronic structure, unstable to the Mott-insulator transition. The application of external pressure reduces the distance between the dimers and increases the width of the conduction band $W$, suppressing the Mott transition. 

At a pressure of $p$ = 5$\,$kbar, Shubnikov-de Haas oscillations with a frequency of $F\approx 240$ T were detected in the quantizing magnetic field in a good agreement with the theoretical calculations of the band structure for the triclinic lattice. A further increase in pressure leads to an increase in the magnitude of magnetoresistance and in the amplitude of quantum oscillations with the same frequency. As a result, it is assumed that localized and delocalized electrons coexist in the entire pressure range $p$ = (1-8)$\,$kbar. Below we show data confirming the spin-glass state in $\kappa$-(BEDT-TTF)$_2$Hg(SCN)$_2$Br at low temperatures and analyze the effect of such a state on the magnetoresistance behavior near the MIT. A strong magnetoresistance at the Mott MIT was investigated before \cite{Brinkman1970,Laloux1994,Rozenberg1994,GeorgesRMP1996,Kagawa2004}, but we argue that our case differs from the standard theory \cite{Laloux1994,Rozenberg1994,GeorgesRMP1996} because of spin frustrations on a triangular lattice and of the strong entanglement between spin state and charge transport. 

In Sec. \ref{SecExperiment} we provide our experimental data on a strong magnetoresistance with a huge hysteresis at pressure around 3$\,$kbar of the Mott transition in a magnetic field $B<16$T. These data show the dependence of this hysteretic magnetoresistance on temperature, pressure and magnetic field. At first glance, the observed strong influence of the rather weak Zeeman splitting $\lesssim 10$K on a Mott MIT with a much larger energy scale $\sim T_{MI}=90$K is very surprising. In Sec. \ref{SecTheory} we explain this interesting effect and illustrate it on a toy model. In Sec. \ref{SecDiscussion} we discuss our results and theoretical model, and indicate some qualitative differences between the proposed and previously known magnetoresistance mechanisms.

\section{Experimental results}
\label{SecExperiment}

Sample resistance was measured using a standard four-probe method by a Lock-in detector on single crystals with characteristic sizes of 1.0x1.0x0.1\,mm$^3$. Two contacts were prepared to each of two opposite sample surfaces with conducting graphite paste. A measuring current with a frequency of 20$\,$Hz, the value of which did not exceed 1$\,\mu$A, was directed perpendicular to the conducting layers of the crystal, so we have measured the out-of-plane sample resistance. The samples were mounted in a high-pressure chamber filled with silicon oil, which ensured operation under quasi-hydrostatic conditions at low temperatures. The pressure was measured by the manganin probe; in all figures we show the value of the pressure corresponding to the low temperature region.  The measurements were carried out in the variable temperature insert that allowed working in the temperature range from the room temperature down to 1.3$\,$K. A magnetic field of up to 7$\,$T was created by a superconducting solenoid and directed either perpendicular or along the conductive layers of the crystal. 

\begin{figure}[tbh]
	\begin{center}
		\includegraphics[width=\columnwidth]{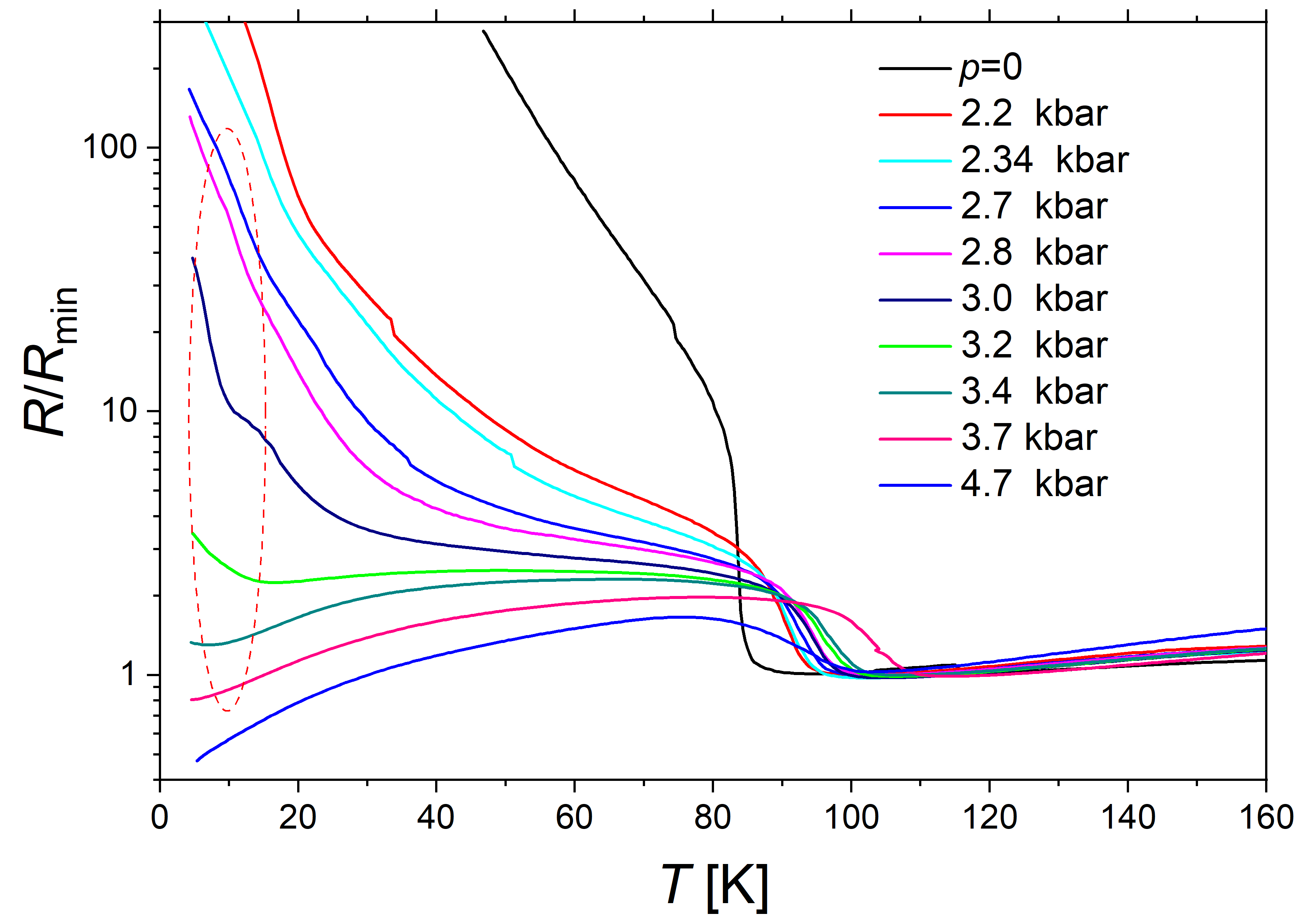}
		\caption{Temperature dependences of the reduced interlayer resistance $R/R_{min}$ of the sample $\kappa$-(BEDT-TTF)$_2$Hg(SCN)$_2$Br at different pressures. $R_{min}$ - the resistance value at minimum on the $R(T)$ dependence in the region of structural phase transition.}
		\label{Fig1}
	\end{center}
\end{figure}

As we have previously established \cite{Pesotskii2022}, in the samples of  $\kappa$-(BEDT-TTF)$_2$Hg(SCN)$_2$Br at $T\approx$90$\,$K a structural phase transformation accompanied by Mott MIT takes place. Fig. \ref{Fig1} shows the temperature dependencies of the interlayer resistance in $\kappa$-(BEDT-TTF)$_2$Hg(SCN)$_2$Br at various pressure values. When the temperature goes down the sample resistance decreases, but at $T\approx$90$\,$K the resistance sharply increases due to the MIT. External pressure reduces the sample resistance and suppresses the MIT without changing the crystal structure. At ambient pressure and room temperature this resistance is $R\approx\,$130$\,$Ohms. At the pressure corresponding to the lowest curve in Fig. \ref{Fig1}, the resistance drops down to 64 $\,$Ohm. All curves shown in Fig. \ref{Fig1} were recorded at a decrease of temperature. Therefore, the hysteresis associated with the structural transition, i.e. shift of transition temperature by about $5\,$K during the up and down temperature sweeps, is not represented here. In the present work, in contrast to Ref. \cite{Pesotskii2022} where the magnetotransport was studied in a metallic state, we focus on the study of low-temperature magnetoresistance in the MIT pressure region 2.5$\,$kbar $< p <4\,$kbar. This area is highlighted in Fig. \ref{Fig1} by a red dotted oval. 

\begin{figure}[tbh]
	\begin{center}
		\includegraphics[width=\columnwidth]{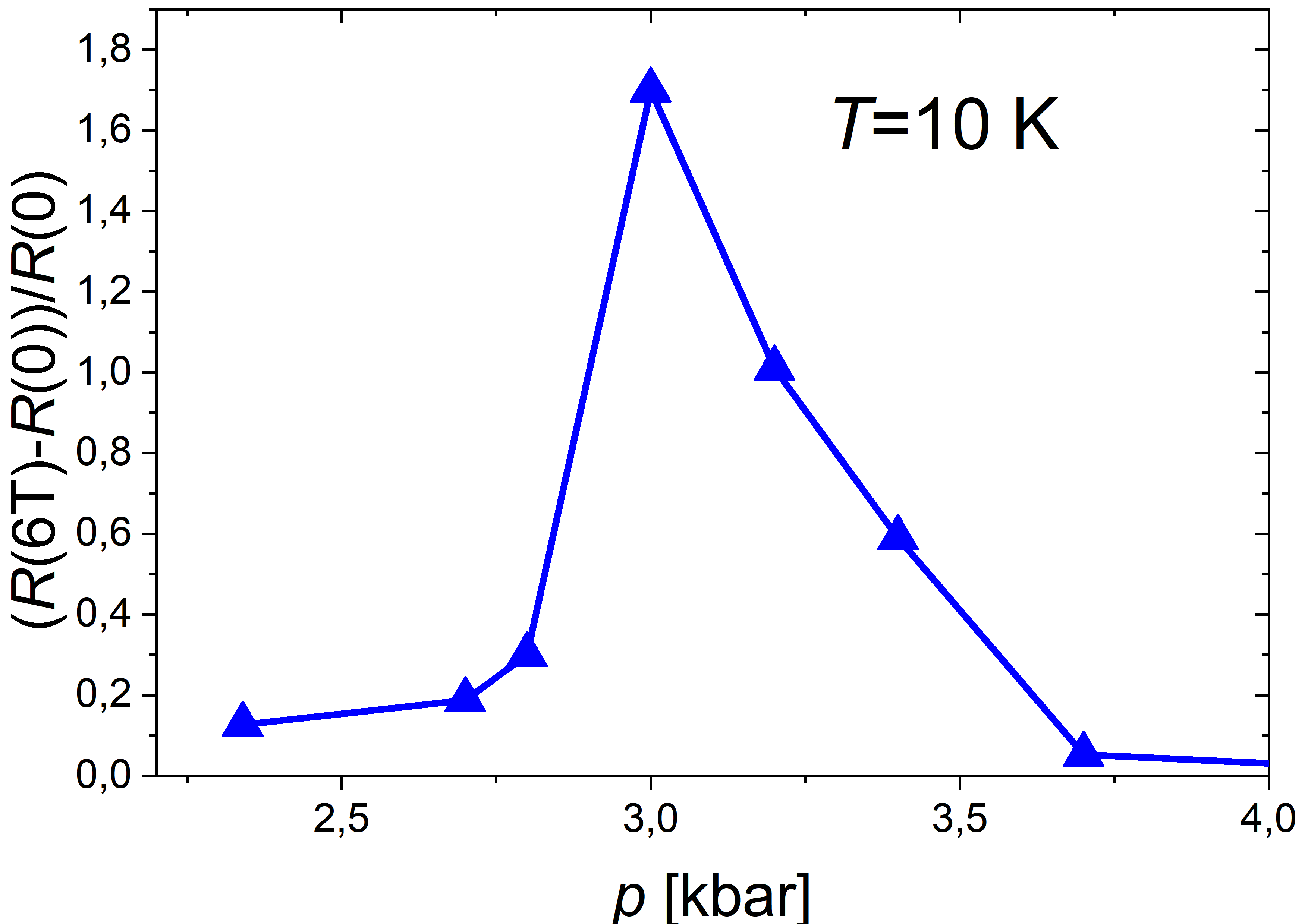}
		\caption{Sample magnetoresistance $(R(6T)-R(0))/R(0)$ at $T$=10$\,$K as function of pressure.}
		\label{Fig2}
	\end{center}
\end{figure}

Fig. \ref{Fig2} shows the pressure dependence of magnetoresistance $\Delta R/R=[R(6T)-R(0)]/R(0)$ at a relatively high fixed temperature $T=\,$10$\,$K. It illustrates an abnormal increase in magnetoresistance in the MIT region, exceeding 150\% at a pressure of about 3 $\,$kbar. Note that at pressures both above and below the transition region, $\Delta R/R$ does not exceed 10\% at the same temperature. 

\begin{figure}[tbh]
	\begin{center}
		\includegraphics[width=1\columnwidth]{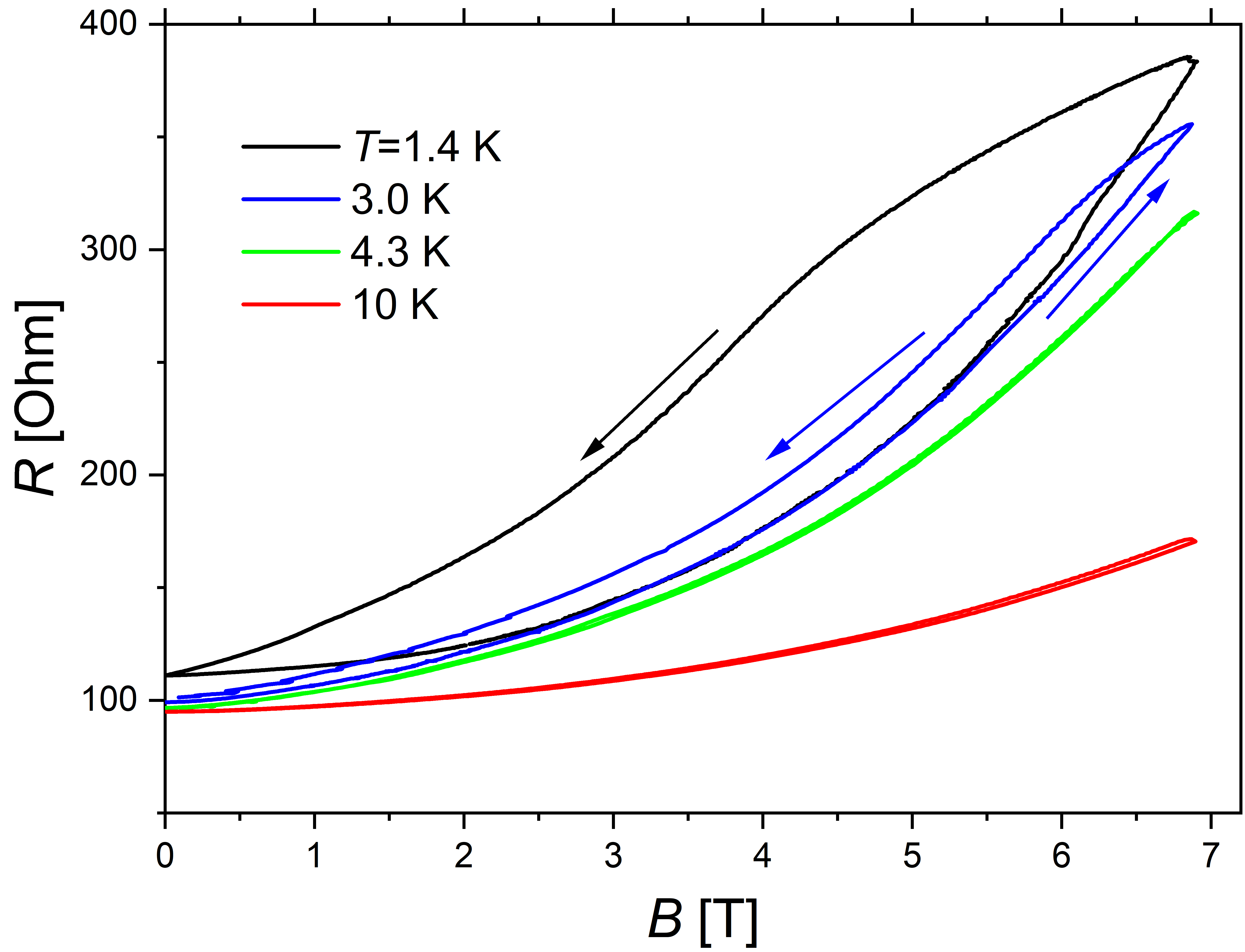}
		\caption{$R(B)$ dependence at various temperatures and pressure $p=3.4\,$kbar.}
		\label{Fig3}
	\end{center}
\end{figure}

\begin{figure}[tbh]
	\begin{center}
		\includegraphics[width=1\columnwidth]{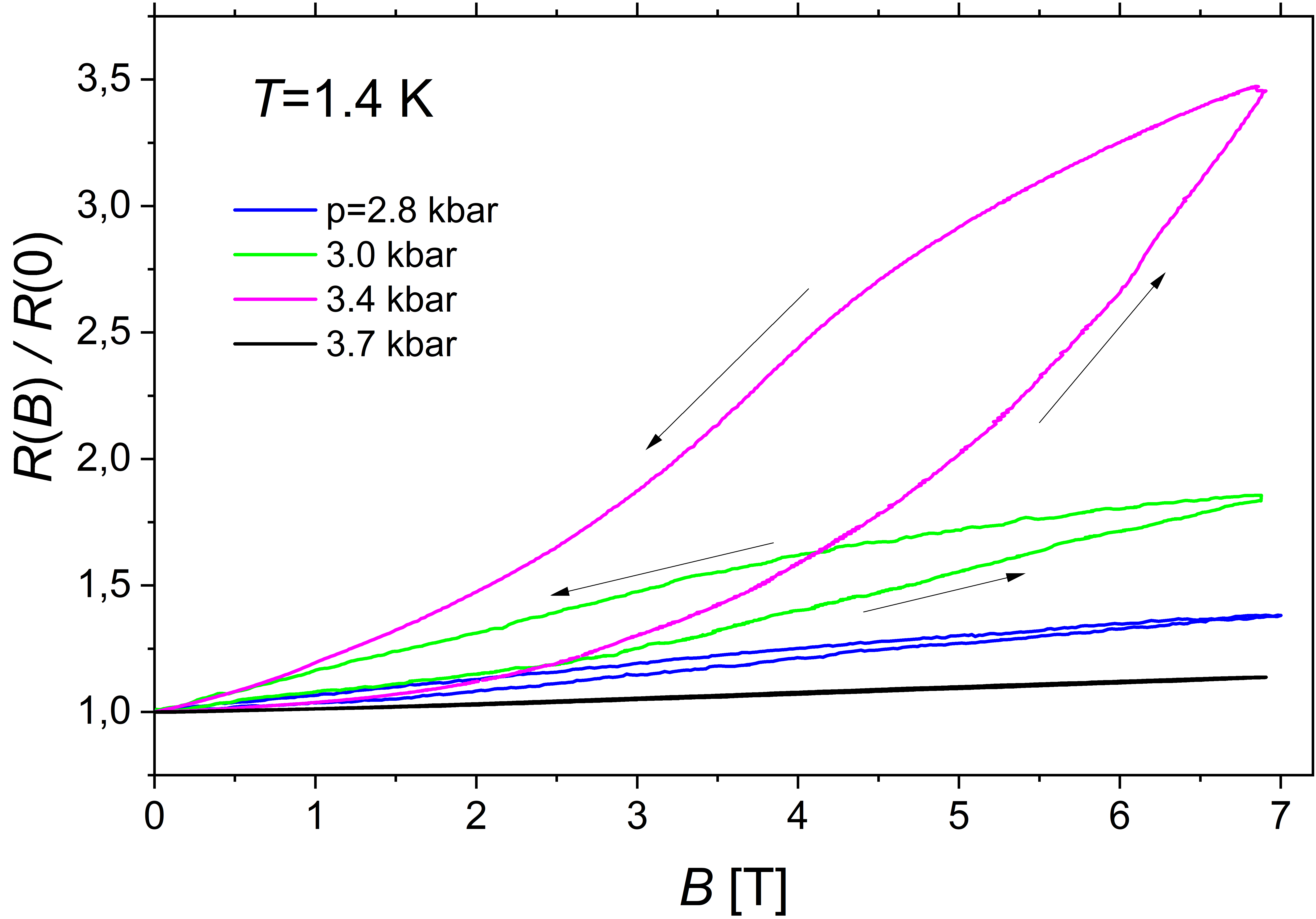}
		\caption{$R(B)$ dependences at various pressures. $T$=1.4$\,$K.}
		\label{Fig4}
	\end{center}
\end{figure}

The most interesting $R(B)$ results were obtained in the low-temperature region $T<10\,$K. Fig. \ref{Fig3} shows these results at pressure $p=3.4\,$kbar. As one can see from this figure, the magnetoresistance increases dramatically with the decrease of temperature, and a strong hysteresis occurs on the $R(B)$ dependencies. Similar curves were obtained at pressures from 2.8 to 3.7$\,$kbar, which is illustrated in Fig. \ref{Fig4}, where the $R(B)$ dependencies are shown for various pressures at $T=1.4\,$K. The hysteresis is present on all curves in Fig. \ref{Fig4}, being almost invisible on the lowest one for $p$=3.7$\,$kbar in this scale, because the maximum loop width on this curve is approximately 1\%, while at $p=3.4\,$kbar the loop width exceeds 50\%.

\begin{figure}[tbh]
	\begin{center}
		\includegraphics[width=1\columnwidth]{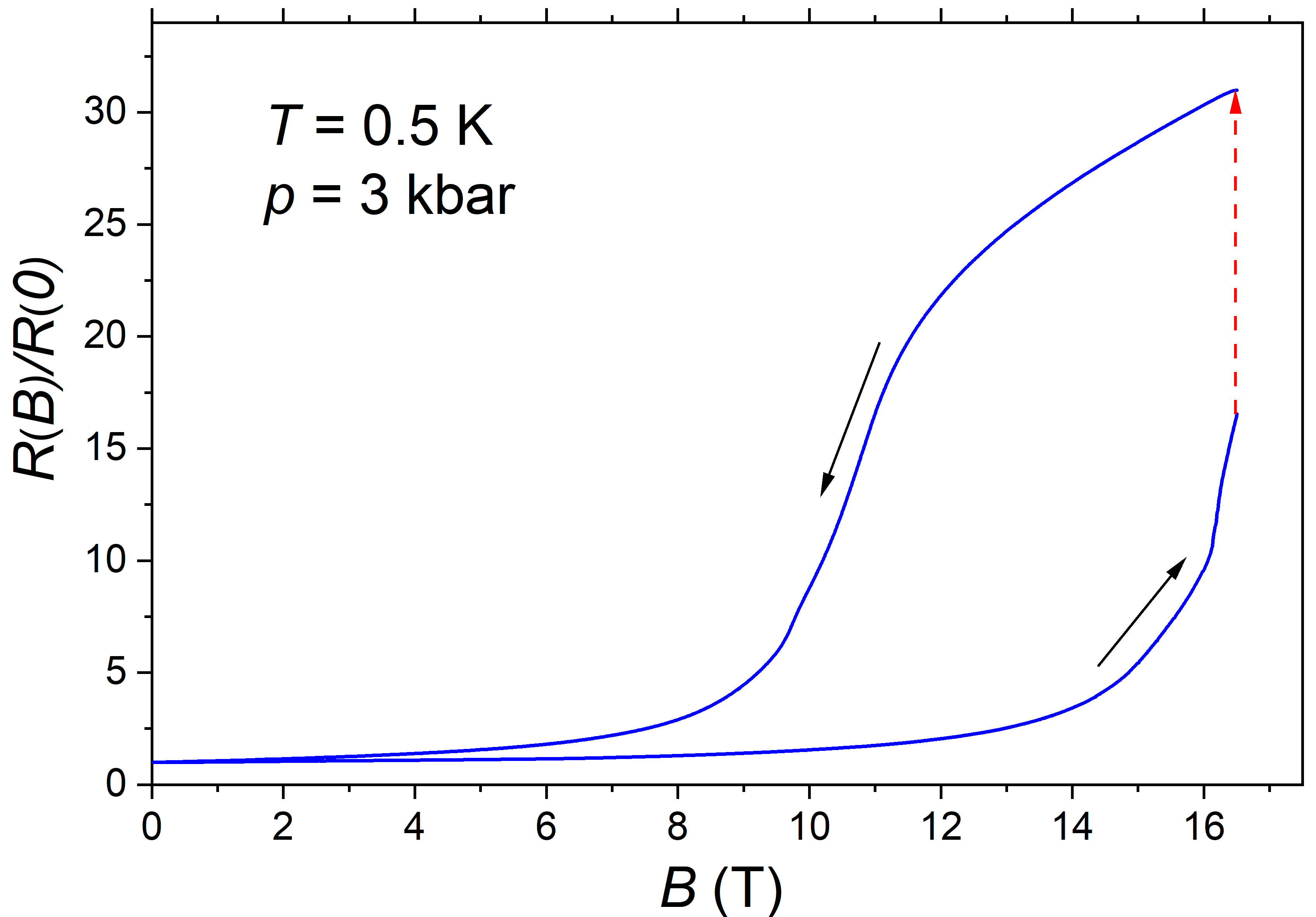}
		\caption{$R(B)$ dependence at $p=3$\,kbar and $T=0.5\,$K.}
		\label{FigHighB}
	\end{center}
\end{figure}

It is noteworthy that the presented in Figs. \ref{Fig3} and \ref{Fig4} $R(B)$ curves do not depend on the magnetic field sweep rate below 0.4$\,T$/min., i.e. with a fixed field $B$ at any point on the curve $R(B)$, the sample resistance practically does not change over time. When the field sweep was stopped, a slight increase in the sample resistance was observed, not exceeding a few percent for several tens of seconds. A different behavior appears at higher fields and lower temperatures. Under these conditions, the effects of long-term resistance relaxation were observed: the sample resistance increased significantly at a fixed magnetic field with a characteristic time of about an hour. Fig. \ref{FigHighB} demonstrates this effect at a pressure 3$\,$kbar in a field of 16.5$\,$T and a temperature of 0.5$\,$K. The resistance growth at a fixed magnetic field is shown by the red vertical arrow. The observation of such a long relaxation time was the main reason for conducting our experiments at higher temperatures and lower magnetic fields. 

\begin{figure}[tbh]
	\begin{center}
		\includegraphics[width=1\columnwidth]{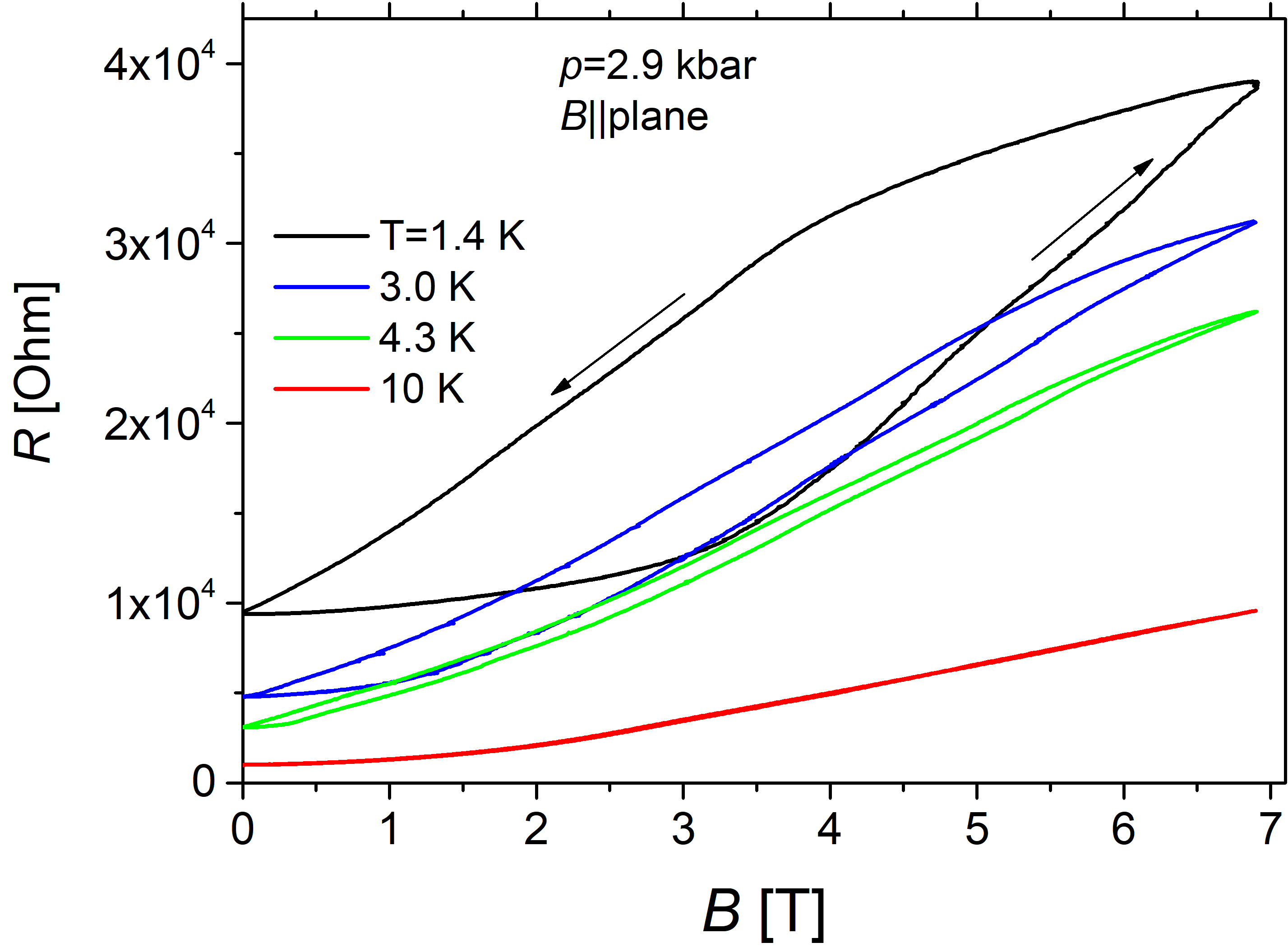}
		\caption{$R(B)$ dependences at various temperatures and fixed pressure $p$=2.9$\,$kbar in the geometry when magnetic field lies in the plane of conducting layers.}
		\label{Fig6}
	\end{center}
\end{figure}

In control experiments, we performed the magnetoresistance measurements in a geometry when the magnetic field was directed along the conducting layers of the sample. It turned out that in this case hysteresis is also observed, and even more pronounced compared to the $B||c$ case. Fig. \ref{Fig6} shows the $R(B)$ dependence at a pressure of 2.9$\,$kbar in the case when the field is parallel to conductive layers. The comparison of these data with those shown in Fig. \ref{Fig4} shows that the observed effects in this geometry are much more pronounced: at $T=1.4\,$ K the hysteresis loop is significantly wider, and the magnetoresistance is several times greater even compared to the curve at $p$=3$\,$ kbar in Fig. \ref{Fig4}.

\begin{figure}[tbh]
	\begin{center}
		\includegraphics[width=1\columnwidth]{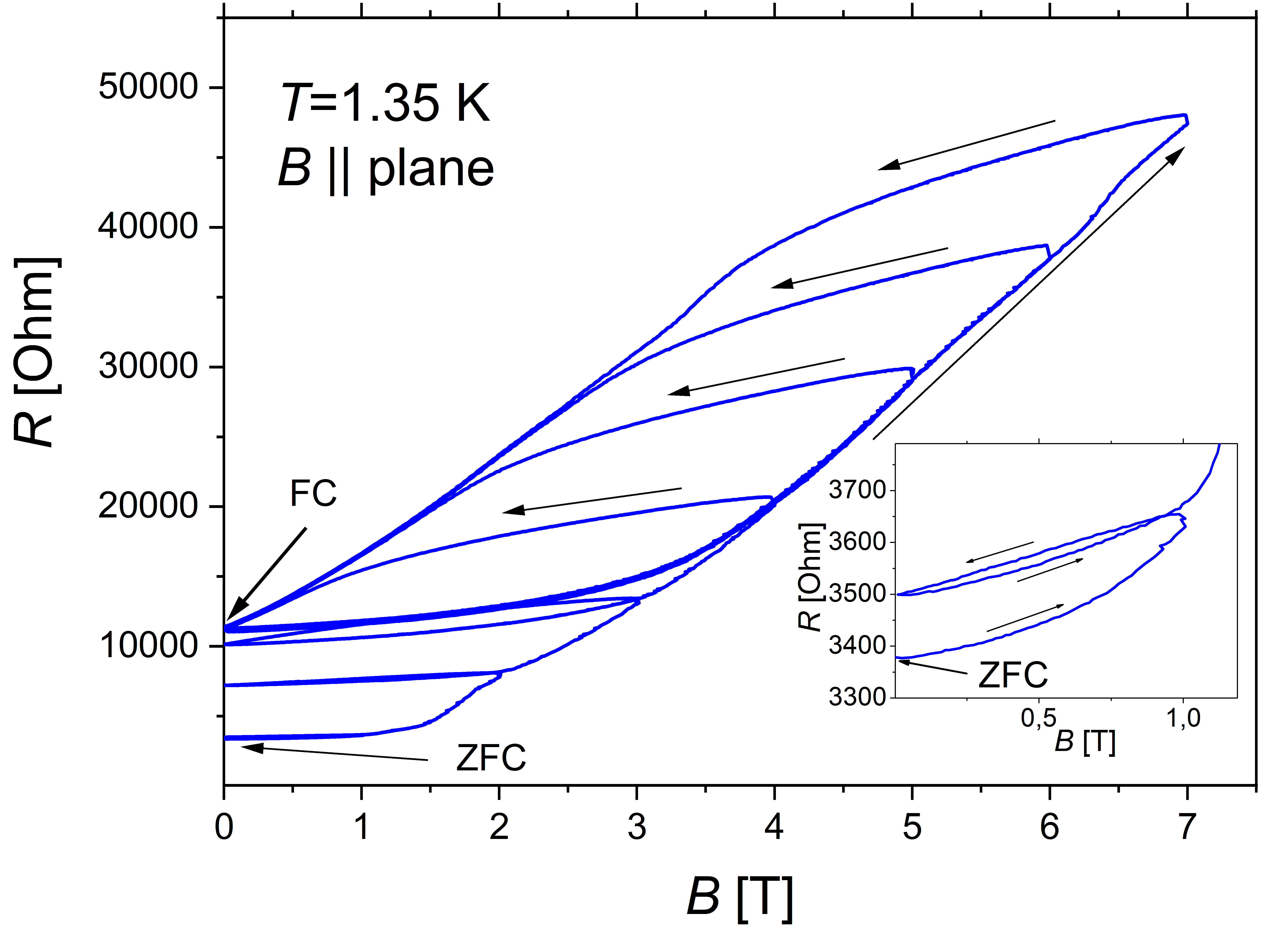}
		\caption{$R(B)$ dependences for cyclic changes of magnetic field from zero to values 1, 2,... 7$\,$T at $T=1.35\,$K. Starting value corresponds to the resistance obtained in field-free cooling mode (ZFC). The insert shows the first cycle on an enlarged scale, which is practically invisible in the main figure.}
		\label{Fig7}
	\end{center}
	
\end{figure}

Since we recorded the influence of the magnetic field on the sample resistance depending on the prehistory, we compared the $R(T)$ dependencies both for zero field cooling (ZFC) and for field cooling (FC) regimes. It turned out that in the ZFC mode, the sample resistance at the final low temperature is several times lower than that in the case of the FC mode. In addition, we determined the value of magnetic field required to transfer the sample to a state corresponding to FC regime. The results of these measurements are shown in Fig. \ref{Fig7} for the case where the field was directed along the planes of conducting layers. In the case where the field was directed perpendicular to the layers, we obtained a similar result. 

The resistance values obtained in ZFC and FC modes are indicated in Fig. \ref{Fig7} by arrows. Starting from the ZFC state, we successively cycled the magnetic field, increasing it to the values 1, 2,... 7$\,$T, each time returning to the point $B$=0. As can be seen from the figure, with a successive increase in the field to 1, 2 and 3$\,$T, the hysteresis loops did not close, and the final resistance was higher than the starting one. However, with an increase in the field to 4$\,$T, the sample resistance returns at $B$=0 directly to the value corresponding to the FC mode, while further cycling the field to larger values already leads to a closed hysteresis loop, i.e. to the FC state.

\begin{figure}[tbh]
	\begin{center}
	\includegraphics[width=1\columnwidth]{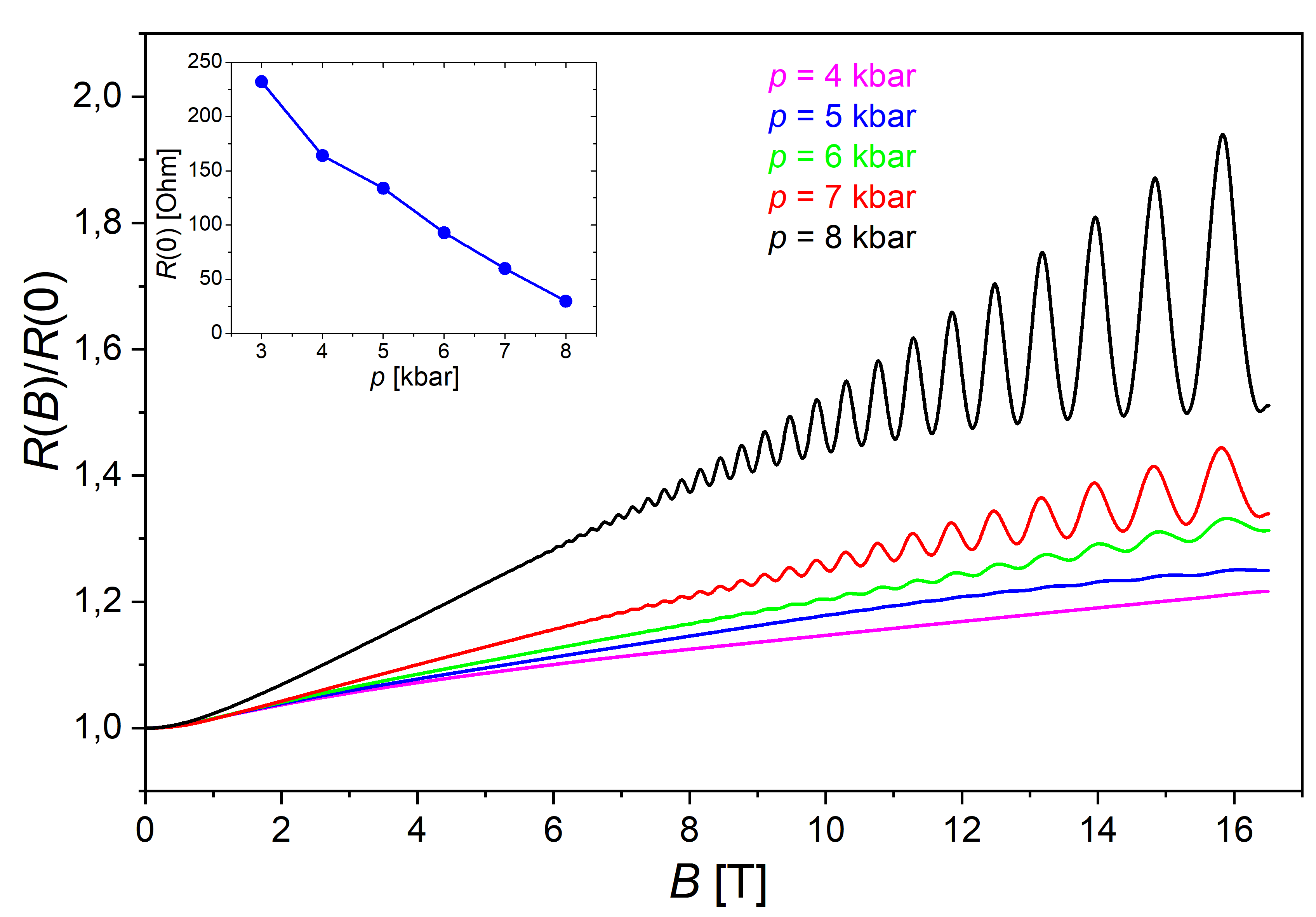}
	\caption{Magnetic quantum oscillations at $T$=0.5$\,$K and various pressures. The pressure dependence of the sample resistance at $B=0$ is shown in the Inset.}
	\label{Fig8}
\end{center}
\end{figure}

As we have argued in \cite{Pesotskii2022} and pointed out in the Introduction, free and localized electrons may coexist, and the fraction of free electrons increases with pressure. This statement is illustrated by Fig. \ref{Fig8}, which demonstrates the low-temperature magnetoresistance in the metallic state at magnetic field perpendicular to conducting layers. One can see that the higher is the pressure, the smaller is the sample resistance (see the insert to Fig. \ref{Fig8}). Moreover, the relative amplitude of the Shubnikov - de Haas oscillations increases dramatically with increasing pressure, while the oscillation frequency does not change. These observations mean that we really deal with the coexistence of insulating and conducting phases, and the fraction of conductive phase increases with pressure.

\section{Theoretical description}
\label{SecTheory}

Let us summarize the main experimental observations and their current understanding. The interdimer electron transfer integrals $t,t'$, which in metals determine the conducting bandwidth, are rather small in $\kappa$-(BEDT-TTF)$_2$Hg(SCN)$_2$Br and comparable to the Coulomb energy. This leads to the Mott metal-insulator transition (MIT) at ambient pressure at $T_{MI}\approx 90$K \cite{3}. Note that at the same temperature the structural phase transition also happens \cite{Pesotskii2022}. This combined structural and electronic phase transition is of the first order, as experimentally  confirmed by the sharp peak of specific heat and by both resistivity \cite{Pesotskii2022} and susceptibility hysteresis at $T\approx T_{MI}$ (see Fig. 5 in Ref. \cite{Hemmida2018}). Hence, the heterogeneous state with both metallic and insulating phases may appear near the MIT. An external pressure $p$ reduces the interdimer distance, increasing the interdimer transfer integrals $t,t'$ and favoring the metallic phase. Hence, the external pressure increases the volume fraction of the metallic phase. At $p>p_c\approx 3\,$kbar the metallic conductivity is observed till very low temperatures, corresponding to the quantum phase transition.  

We observe a giant magnetoresistance in $\kappa$-(BEDT-TTF)$_2$Hg(SCN)$_2$Br at the Mott transition, see Fig. \ref{Fig2}, which becomes much stronger at low temperature (see Figs. \ref{Fig3}-\ref{Fig8}). This effect is not new in organic metals and was previously investigated in $\kappa$-(BEDT-TTF)$_2$Cu[N(CN)$_2$]Cl \cite{Kagawa2004}. 
The common theoretical description of this magnetoresistance is based on the following. The magnetic susceptibility $\chi $ near the Mott transition diverges in the Gutzwiller's approximation \cite{Brinkman1970} or strongly increases according to the numerical calculations based on the Hubbard model at half filling in the controlled limit of infinite spatial dimensionality \cite{Laloux1994,Rozenberg1994}, which is actually a dynamic mean field theory (DMFT) method \cite{GeorgesRMP1996}. Hence, the external magnetic field $\boldsymbol{B}$ may considerable influence the Mott phase transition and shift the transition line, thereby giving rise to magnetic-field-induced Mott transition, or magnetoresistance. When $\boldsymbol{B}$ shifts the system toward the Mott-insulating phase, it increases the electric resistivity, leading to positive magnetoresistance. This explanation is adopted, e.g., in Ref. \cite{Kagawa2004}.

However, our case have several important differences from the usual case, described by the standard theory \cite{Laloux1994,Rozenberg1994,GeorgesRMP1996}. 
First, we observe a huge hysteresis in $\kappa$-(BEDT-TTF)$_2$Hg(SCN)$_2$Br, and it happens at temperatures much below the Mott MIT but close to the spin-glass transition temperature. As seen in the Fig. \ref{FigHighB}, the magnetoresistance value $(R(16.5$\,T)-$R(0))/R(0)$ at 0.5\,K  is about 30, (i.e., 3000\% !), but already at $T\sim 10$\,K, which is still much lower than the Mott-insulator transition temperature $T_{MI}\approx 90$\,K, magnetoresistance is by an order of magnitude smaller (see Fig. \ref{Fig3}). Note that the hysteresis at $T_{MI}=90\,$K, associated with the first-order Mott phase transition as described by standard theory \cite{Laloux1994,Rozenberg1994,GeorgesRMP1996}, is also present (see Fig.  5b of Ref. \cite{Hemmida2018} and Fig. 2b of our previous paper \cite{Pesotskii2022}), but this hysteresis has a narrow temperature range around $T_{MI}=90\,$K and closes at $T<87.5\,$K, indicating that it is completely different from the hysteresis at $T<10\,$K studied in the present paper. 

Both the magnetoresistance and the width of hysteresis loop increase with the field and with the drop of temperature. The observed dependence of magnetoresistance on prehistory (see Fig. \ref{Fig7}) is obviously related with the influence of frustrated spin system and of the corresponding spin-glass state on the measured resistance. From Fig. \ref{Fig7} one sees that the hysteresis loops close if the field $B\ge$ 4\,T, when the Zeeman energy $E_Z\sim 5\,$K is much greater than $T=1.35$. The characteristic temperature when the strong hysteresis appears has a close magnitude $T_{hyst}\sim 5$\,K (see Fig. \ref{Fig6}). This temperature $T_{hyst}\sim E_Z$ is much lower than the Mott-insulator transition temperature $T_{MI}\approx 90$\,K, indicating that the observed hysteresis is not determined by the Mott-insulator transition only, as implied in the standard theory \cite{Laloux1994,Rozenberg1994,GeorgesRMP1996}. The hysteresis energy scale $T_{hyst}\sim E_Z\sim 5$K is very close to the temperature $T_m\approx 5$\,K of excess peak in heat capacity, attributed to the glass behavior \cite{Hemmida2018}. On the other hand, the entropy contained in the excess peak $S = 13.7$ J/mol\,K is significantly larger than the pure magnetic entropy of the spin-1/2 system ($S = R \ln 2$), indicating that the glassy state largely involves the charge degree of freedom, resulting in a charge-spin entanglement \cite{Hemmida2018}. Our model below qualitatively describes the physics of this entanglement.

Second, the observed very slow relaxation of magnetoresistance (see Fig. \ref{FigHighB}) also unambiguously indicates the strong influence of spin-glass state, having this slow dynamics (see Supplemental Materials), on magnetoresistance. The energies of states with opposite spins in the field differ by $\Delta E_Z$, so the change in the configuration of the spin system occurs over a time that depends on the field and temperature: it increases with decreasing $T$ and increasing $B$. 

Third, even with the enhanced susceptibility near the Mott phase transition, the simple thermodynamic effect of magnetic field is, probably, not sufficient to strongly shift the balance between Mott-insulating and metallic phases because of a large difference of energy scales. The magnetic field $B\leq 7\,$T even at saturation of magnetization introduces the energy scale of Zeeman energy $E_Z \sim \mu_B B < 10\,$K per electron, while the Mott MIT temperature is $90\,$K, and the corresponding activation energy $E_g\sim 20$\,meV $\approx 230\,$K \cite{3} is even higher. The formation of the energy gap $E_g$ at the Fermi level with density of electron states $\nu_F$ gives the energy gain $\sim E_g^2 \nu_F /2 $ of Mott-insulating phase, which considerably exceeds the magnetic energy gain $\chi B^2/2$. This comparison is easier to perform per one electron. The energy gain from the formation of Mott energy gap at the Fermi level is $\Delta E_{MI}\sim E_g^2 /2E_F$ per one electron \footnote[1]{This estimate has a simple physical meaning: due to lowering of the energy of filled electron states by the formation of energy gap, each electron at the Fermi level gains the energy $\sim E_g/2$. The fraction of such electrons at the Fermi level is $\nu \sim E_g/E_F$. Hence, the total energy gain per electron is $\sim \nu E_g\sim  E_g/E_F$.}, where $E_F$ is the Fermi energy in the metallic phase. In $\kappa$-(BEDT-TTF)$_2$Hg(SCN)$_2$Br the band structure calculations give \cite{Pesotskii2022} $W\approx 0.2\,$eV, and the corresponding Fermi energy at half filling is $E_F\sim 0.1\,$eV. Substituting the numbers, we get the energy gain $\Delta E_{MI}\sim E_g^2 /2E_F\approx 0.002\,$eV$\sim 25\,$K per electron in conducting band, which considerably exceeds the upper limit of magnetic energy gain $E_Z$ from these electrons. Indeed, even at complete spin polarization of conducting band, the magnetic energy gain is $E_{mag}< E_Z\approx \mu_B B< 8\,$K per one electron at field $B\leq 7\,$T. Due to the antiferromagnetic coupling on a triangular lattice, a considerable fraction of spins is not aligned along the magnetic field, which additionally reduces $E_{mag}$ by a factor $\sim 3$. In a metallic state the magnetic energy gain from conducting electrons $E_{mag}$ is additionally reduced by a factor $\sim E_Z/E_F<0.01$. Hence, even with the enhanced susceptibility near the Mott phase transition, the simple thermodynamic effect of magnetic field is not sufficient to strongly shift the Mott transition line. Below we argue that the effect of magnetic field in $\kappa$-(BEDT-TTF)$_2$Hg(SCN)$_2$Br is not reduced to simple thermodynamics, because the spin alignment directly affects the electron transport.

Although the magnetic field evidently shifts the Mott transition line according the standard DMFT theory \cite{Brinkman1970,Laloux1994,Rozenberg1994,GeorgesRMP1996}, we believe that this theory is not sufficient to explain all observed features of the Mott transition in $\kappa$-(BEDT-TTF)$_2$Hg(SCN)$_2$Br. The spin frustrations on a triangular lattice must be included into the model as they play an essential role in physical processes. This is especially reflected in slow relaxation and giant hysteresis, common for spin-glass state. The Hubbard model on a triangular lattice has received a considerable theoretical attention \cite{Aryanpour2006,PhysRevB.96.205130,PhysRevB.102.115142,PhysRevX.11.041013,nano11051181}, but the effect of external magnetic field on this system was not sufficiently investigated. There may be additional difficulties of applying the standard DMFT methods with a local self-energy part to the spin-glass system where the non-local effects are essential. Below we do not perform novel DMFT calculations for the half-filled Hubbard model on the triangular lattice in a magnetic field, but propose a simple qualitative model which, in our opinion, captures the essential physics of the observed giant magnetoresistance with a huge hysteresis and slow relaxation at low temperature. It also helps to understand the physical mechanisms behind the complicated ab initio calculations.

\subsection{Qualitative description}

To present our qualitative idea behind the proposed colossal magnetoresistance mechanism and its hysteretic behavior we first emphasize several features of $\kappa$-(BEDT-TTF)$_2$Hg(SCN)$_2$Br. The conducting layers of the dimerized BEDT-TTF molecules hosts one electron per dimer. The dimer size is $d\sim 1$\,nm, and the dimensional quantization on this small size gives a large energy splitting $\Delta E_d \sim \hbar^2 /(md^2) \sim 0.1$\,eV$\gg T\sim 2$\,K. The electrons mainly occupy only the lowest energy level on each dimer. The spins of these electrons couple antiferromagnetically with similar electrons on neighboring dimers \cite{Hemmida2018}. The opposite spin orientations of neighboring electrons allow interdimer jumps which delocalize the electron wave functions, thus lowering the electron kinetic energy. It also enhances the electron conductivity and favors the metallic behavior. For ferromagnetic spin orientation on adjacent dimers the interdimer jumps is forbidden by Pauli exclusion principle. We call such a ferromagnetically-ordered spin configuration of electron pair on adjacent dimers as \emph{current-locking}. 

The triangular lattice of BEDT-TTF dimers leads to geometrical magnetic frustrations and spin-glass behavior instead of long-range AFM order \cite{Hemmida2018}. The observed excess peak in the specific heat at $T=T_m\approx 5\,$K and the strong increase of magnetic susceptibility at $T<30\,$K  (see Fig. 5 in Ref. \cite{Hemmida2018}) can be attributed to magnetic frustrations and even result to weak ferromagnetism in the form of isolated small ferromagnetic puddles \cite{Hemmida2018}. The magnetic frustrations are also consistent with the observed inhomogeneous NMR line broadening at $T<40\,$K and with the strong increase of the nuclear spin relaxation rates $1/T_1$ and $1/T_2$ at $T<20\,$K \cite{Li2020}. Because of the frustrations, some neighboring electron spins have the same projection, which forbids interdimer hopping between them and reduces conductivity. The typical energy difference between two opposite spin projections is also reduced by the frustrations and is of the order of the temperature of specific-heat maximum $T_m\approx 5$\,K. 

A moderate magnetic field $B\gtrsim 1$T, resulting to Zeeman energy splitting $\Delta E_Z\gtrsim 2$K$\sim T_m$, may strongly change the spin-glass state by aligning a considerable part of electrons on dimers along the magnetic field. This strong response of spin glass in $\kappa$-(BEDT-TTF)$_2$Hg(SCN)$_2$Br is supported by the strong enhancement at $T<20\,$K and even the divergence of magnetic susceptibility at $T\to 0$, as shown in Fig. 4b of Ref. \cite{Hemmida2018}. This increases the number of current-locking ferromagnetically ordered adjacent electron spins in a magnetic field. Hence, it shifts the Mott MIT toward the insulating phase, which results to a strong magnetoresistance in the vicinity of MIT. 

All above referred to the spin system in the dielectric regions of the sample. However, we emphasize that we measure the resistance of the sample using it as a sensor of the spin system's state, i.e., the resistance clearly reflects the state of the spin glass. Therefore, the observed effects (giant magnetic resistance, hysteresis, and temporal relaxation) in combination with clear magnetic quantum oscillations (MQO) can be explained by the contribution of both metallic and Mott-insulating regions to the measured resistance when the current passes sequentially through these regions.

The magnetoresistance hysteresis appears due to the hysteretic behavior of spin-glass configuration with magnetic field variations. For each value of external magnetic field the spin-glass system has a new set of local energy minima and of the corresponding spin configurations. The system chooses a local energy minimum, which is far from the global minimum. The relaxation of this spin glass system is rather slow. In our experiment at $B = 16.5 \,$T and $T = 0.5 \,$K (see Fig.5.), when $\Delta E_Z\gg T$, the typical relaxation time of the resistivity was about one hour, but in a weaker magnetic field and higher temperature the relaxation time is much smaller. This strong magnetic-field dependence of relaxation time indicates that the spin glass is involved. Hence, a typical glassy properties of the relaxation time observed in $\kappa$-(BEDT-TTF)$_2$X family by fluctuation spectroscopy studies \cite{CrystReview2018} is related not only to charge but also to spin dynamics. When the magnetic field increases, the system chooses a local minimum which is closer to its zero-field configuration, where the number of AFM spin bonds is maximal. This favors the metallic phase. When the magnetic field decreases, the system chooses a local minimum which is closer to its high-field almost spin-polarized configuration, where the number of current-locking ferromagnetic spin bonds is maximal, favoring the insulating phase. Hence, at the same conditions, the resistivity in a raising magnetic field is smaller than in a decreasing field.   

The above idea describes qualitatively both the giant magnetoresistance and its hysteresis at the Mott-Hubbard phase transition, as we observed in $\kappa$-(BEDT-TTF)$_2$Hg(SCN)$_2$Br at pressure $\sim $3$\,$kbar. A quantitative theory of this unusual magnetoresistance behavior, originating from strong electronic correlations, requires at least complex and lengthy numerical calculations, e.g. using the dynamical mean-field theory (DMFT) (see, e.g., \cite{KotliarDMFT2006,Lyakhova2023} for reviews). This description is additionally complicated by the presence of magnetic field, spin-frustrations and non-local spin-glass behavior, and by subsequent conductivity calculations. While DMFT often gives a reasonable band structure at the Mott-Hubbard phase transition on a large energy scale $\gtrsim 0.1$\,eV \cite{KotliarDMFT2006,Lyakhova2023}, calculating electron transport properties using the derived electronic structure, especially in a magnetic field, requires precise details that are not typically captured by DMFT. The available DMFT calculations of the Hubbard model at half filling in a magnetic field \cite{Laloux1994,Rozenberg1994,GeorgesRMP1996} do not take into account the spin frustrations and glassy behavior on a triangular lattice. Therefore, to illustrate that the observed giant hysteretic magnetoresistance indeed originates from the above qualitative arguments, and that the spin frustrations and spin-glass behavior strongly affect the magnetoresistance, we consider a simple classical toy model, as described below. 

\subsection{Toy model}

\begin{figure}[tbh]
	\begin{center}
		\includegraphics[width=0.6\columnwidth]{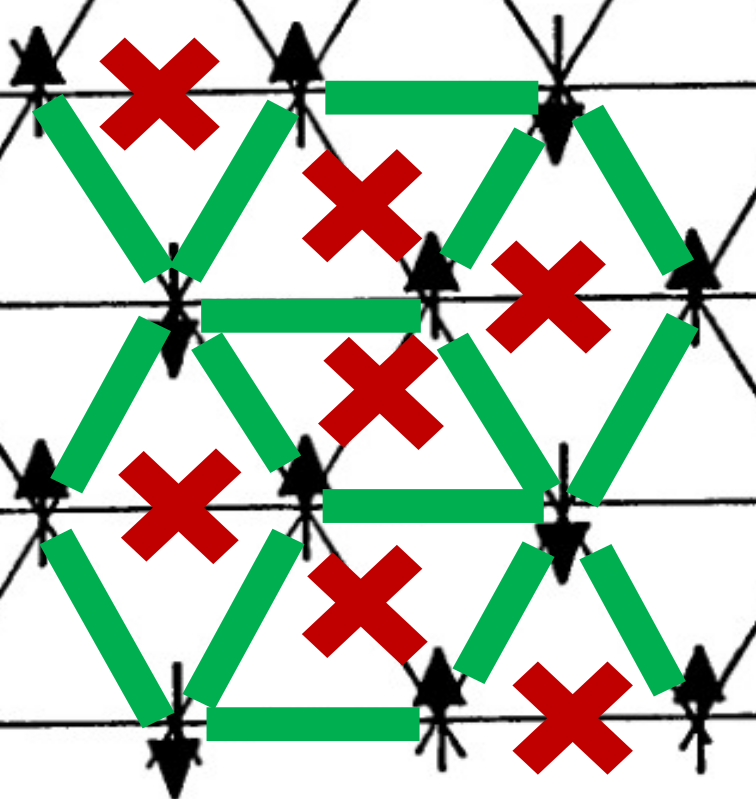}
		\caption{Schematic illustration of the classical toy model of a triangular lattice with one electron per site on the Fermi level. The electrons may hop to the nearest sites, but the probability of this hopping depends very much on the mutual spin orientation of the electrons on these two nodes. Therefore, the nearest sites of this triangular lattice are connected by the resistances of two values $R_1$ (green thick lines, when the electron spins are antiparallel) and $R_2\gg R_1$ (red crosses), corresponding to the same spin orientation on these nearest sites.}
		\label{FigScheme}
	\end{center}	
\end{figure}

\begin{figure}[tbh]
	\begin{center}
		\includegraphics[width=\columnwidth]{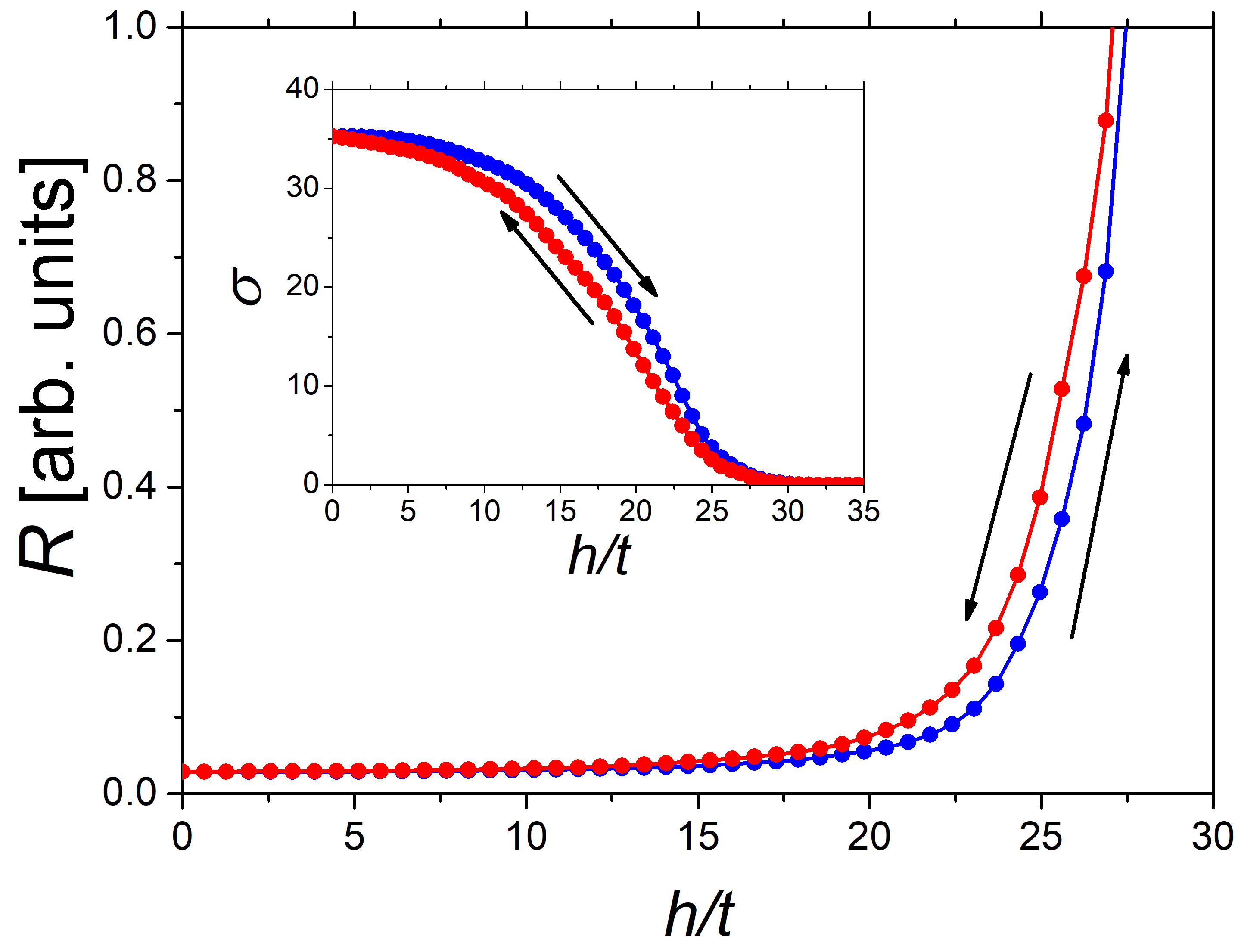}
		\caption{Calculated giant magnetoresistance $R(B)$ in our toy model for increasing (blue symbols) and decreasing (red symbols) magnetic field. The magnetic field $h/T\propto B$ is normalized as the ratio of Zeeman energy $h=\Delta E_Z/2$ to temperature $T$. Inset shows the conductivity $\sigma(B)=R^{-1}$ for the same parameters, where the hysteresis is clearer seen.}
		\label{FigRtheory}
	\end{center}
\end{figure}

Consider a 2D triangular lattice of spins up or down (see Fig. \ref{FigScheme}), which models the conducting electrons on the molecular dimers in $\kappa$-(BEDT-TTF)$_2$Hg(SCN)$_2$Br in one conducting layer. Most time each node of this triangular lattice is occupied by one conducting electron, because two electrons on one dimer cost an extra Coulomb energy.  Nevertheless, the electrons may hop from one node to a nearest node, but the probability of this hopping depends very much on the mutual spin orientation of the electrons on these two nodes: the hopping to the same orbital quantum state is prohibited by the Pauli exclusion principle if two electrons have the same spin projection and is allowed for different spin orientations. Therefore, the nearest nodes of triangular lattice in our toy model are connected by the resistances of two values $R_1$ (green thick lines in Fig. \ref{FigScheme}) and $R_2$ (red crosses in Fig. \ref{FigScheme}). The largest resistance $R_2\gg R_1$ connects the nearest neighbours with the same spin projection of conducting electrons, when the intersite hopping is almost prohibited. Hence, the Ising system of spins on a triangular lattice, easily controlled by external magnetic field, affects strongly the electron conductivity and produces magnetoresistance.

The phase diagram of a classical 2D Ising model on a triangular lattice is well known \cite{SaitoIgeta1984}, but we are interested in the resistance of the lattice affected by its spin configuration and driven by the external magnetic field. Somewhat similar resistive Ising model was studied recently \cite{BarrowsPRE2025}, where the random-graph nanowire network tunnel junctions also take binary conductance values depending on mutual magnetic orientation. However, the different geometry in Ref. \cite{BarrowsPRE2025} gives completely different phase diagram and resistance behavior, not applicable to our system. The memristive Ising circuits \cite{Pershin2022} also have no direct relevance to our system but may give its further prospects.

In Fig. \ref{FigRtheory} we show the results of our numerical calculation of resistivity and conductivity in the proposed toy model as a function of increasing (blue symbols) and decreasing (red symbols) magnetic field. One clearly sees both a giant magnetoresistance and a considerable hysteresis near the metal-insulator transition, as observed in our experiment. In the Supplementary materials we give the details of this calculation, including the dependence of the results obtained on various parameters: temperature $T$, exchange magnetic coupling $J$ of conducting electrons on the closest dimers, the resistance ratio $R_2/R_1$, the speed of the variation of magnetic field, and on the disorder in exchange coupling.

 \section{Discussion}
 \label{SecDiscussion}

The proposed classical toy model describes well the giant magnetoresistance, as shown in Fig. \ref{FigRtheory}. In addition it gives a hysteresis, which is better seen on the conductivity plot in the inset in Fig. \ref{FigRtheory}. It also gives the observed glassy properties of relaxation times (see Supplemental Materials, Figs. 13-17).
However, to obtain a hysteresis in our model one has to introduce a disorder to the frustrated spin system, e.g. in the exchange coupling constants. 
This observation, that the disorder is necessary for hysteresis in addition to a geometrical frustration, is not the peculiarity of our model and agrees with the known general theory of frustrated spin systems \cite{Dotsenko1993,DotsenkoBook}. There are several reasons for the disorder in organic conductors. First, the orientational disorder in conformation of ethylene groups (eclipsed or staggered) of terminal ethylene groups of ET molecules may exist \cite{Leung1985,Schultz1986,Hartmann2014,Guterding2015}. Second, the disorder may appear due to the charge disproportionation on ET molecules \cite{Liebman2024}. 
 Third, the disorder may also originate from electron dynamics. Indeed, the exchange coupling strongly depends on the overlap of electron wave functions. Due to the Coulomb interaction and strong electron correlations, the electron wave functions on each dimer depend on other electrons and on their spins, which may (slowly in comparison to fast electron dynamics) change with time due to spin relaxation, especially when the magnetic field varies. Therefore, the proposed toy model of classical resistances does not give a quantitative agreement of the hysteresis width, because it does not solve the complicated problem of strong electronic correlations properly. In particular, we consider only the effect of external magnetic field on the frustrated electron spins, which affects electronic transport. However, the spin-dependent interaction between conducting electrons does not include the back influence of electron transport and dynamics on their spin configuration.
 
 The proposed toy model is 2D, while in experiment we have a layered 3D material. One can generalize the numerical calculations to a 3D case but the result will be qualitatively the same while it requires a much larger calculation time. Indeed, similar to the intralayer 2D transport, the interlayer hopping of conducting electrons between the adjacent dimers will be hindered or nearly prohibited by the Pauli principle if the electron spins are the same, and allowed for opposite spins. The external magnetic field due to the Zeeman energy gain orients more electron spins along the magnetic field, thus increasing the number of forbidden hopping directions and the total resistivity both along and across the conducting layers. 
In addition, there is a general rule known from various MIT driven by disorder \cite{Lee1985} that if electrons in a metal become more localized in some directions this also reduce conductivity and often produce localization along all directions.
 
In the current classification of different GMR mechanisms, the proposed GMR mechanism is closer to extreme magnetoresistance (XMR) \cite{Niu2022}, because it originates neither from magnetic layers of different magnetization, as in usual GMR \cite{NobelLectureGMR2008,Ennen2016}, nor from a ferromagnetic-to-paramagnetic phase transition inherent to CMR \cite{Ramirez1997,Tafra2025,Balguri2025}. In both these standard mechanisms the crucial role is played by the electron scattering on the macroscopic boundary between two areas of different magnetization, either between two layers of different material and magnetization (GMR) or between ferromagnetic and paramagnetic phases (CMR), which also have different magnetization. In our mechanism there is no scattering by any macroscopic boundary, necessary for usual GMR and CMR mechanisms.   
 
 Contrary to most GMR, our mechanism is almost isotropic both to the direction of magnetic field and electric current. This isotropy to the direction of magnetic field is clear from the comparison of Figs. \ref{Fig6} or \ref{Fig7}, where the magnetic field $\boldsymbol{B}$ is applied parallel to conducting x-y planes, with other figures, e.g. Figs. \ref{Fig3}-\ref{FigHighB} where the magnetic field $\boldsymbol{B}||z$ is normal to conducting layers. This isotropy excludes various orbital magnetoresistance mechanisms common to XMR \cite{Niu2022}. Besides the classical magnetoresistance \cite{Abrikosov1988Fundamentals,Ziman1972Principles},
 there are many other orbital mechanisms of strong interlayer magnetoresistance, pertinent to very anisotropic metals in a field perpendicular to conducting layers \cite{Gvozdikov2007,Kartsovnik2009Apr,Grigoriev2011Jun,Grigoriev2011Sep,Grigoriev2012Oct,Grigoriev2013Aug,Sinchenko2017}. These mechanisms, although being present, are not the main source of the observed GMR in $\kappa$-(BEDT-TTF)$_2$Hg(SCN)$_2$Br because of (i) its nearly isotropy to the direction of magnetic field and (ii) the strong non-monotonic pressure dependence with the maximum at Mott MIT.

At first glance, the observed strong influence of a rather weak Zeeman splitting $E_{Z}\approx eB\hbar /m_e c \lesssim 10$\,K at field $B\lesssim 10$\,T on a MIT with a much larger energy scale $\gtrsim T_{MI}=90$\,K is very surprising. The band width and e-e interaction competing in this Mott transition have even larger energy scale. However, even a much smaller field $B\sim 1\,T$ at field-cooling regime already gives a considerable effect, as shown in Fig. \ref{Fig7}. This puzzling behavior is explained by our model where even a small Zeeman splitting strongly affects the spin-glass state by aligning a considerable part of electron spins along the magnetic field, as supported by both the theory of spin-glass behavior \cite{Dotsenko1993,DotsenkoBook} and by the observed strong enhancement of magnetic susceptibility in $\kappa$-(BEDT-TTF)$_2$Hg(SCN)$_2$Br at low temperature \cite{Hemmida2018}. It is the Pauli exclusion principle rather than the small Zeeman splitting which prevents electron transport in our GMR mechanism.

In our system, the Mott phase transition spans a remarkably wide pressure interval. Throughout the entire pressure range of our experiment, both metallic and Mott‑insulating phases coexist. The insulating fraction is evidenced by the temperature‑dependent resistivity with insulating behavior at  $T<T_{MI}\approx 90$\,K (see Fig. \ref{Fig1}). The metallic fraction is revealed via magnetic quantum oscillations (see Fig. \ref{Fig8}), with metallic domains exceeding the Larmor radius in size. Usually, an external pressure drives the Mott transition much faster, especially in organic conductors which have a rather small elastic modulus. The extended pressure interval of phase coexistence may arise from glassy behavior coupled with the Mott transition. Notably, a strongly enhanced magnetoresistance is in a narrower pressure interval of $\sim 1$ kbar (see Fig. \ref{Fig2}).

Quantifying the pressure‑dependent volume fractions of metallic and insulating phases would provide valuable insight. The simplest approach employs resistivity measurements combined with effective‑medium theory for heterogeneous media \cite{Torquato2002}, which can be easily generalized for quasi-2D layered \cite{Sinchenko2017SC} or completely anisotropic conductors \cite{Seidov2018}. However, this method requires the knowledge of resistivity in the homogeneous metallic and Mott‑insulating phases -- values that remain unknown in our case. The insert to Fig. \ref{Fig8} shows that the dependence $R(p)$ does not saturate at high pressure, even at $p=8$ kbar. This means that even at our maximum pressures both metallic and dielectric phases persist, precluding determination of the metallic‑phase conductivity. Similarly, even at the lowest (ambient) pressure the resistivity $R(p)$ does not saturate \cite{Pesotskii2022}. Consequently, resistivity measurements alone cannot reliably yield phase fractions as a function of pressure. 

Some insight on the portion of metallic phase can be obtained by analyzing the amplitude of magnetic quantum oscillations as a function of pressure. However, the measured Shubnikov-de Haas effect is not very convenient for quantitative estimates, as resistivity depends not only on volume fractions of two phases but also on their morphology (shape and mutual orientation) \cite{Torquato2002,Sinchenko2017SC,Seidov2018}. 
The de Haas-van Alphen oscillation amplitude offers a more suitable probe for volume fraction estimation, particularly when domain sizes much exceed the cyclotron radius of conducting electrons. Nevertheless, the unknown MQO amplitude in the pure metallic phase remains a critical limitation, preventing accurate determination of the metallic-phase fraction.

In another organic conductor $\kappa$-(BEDT-TTF)$_2$Cu$_2$(CN)$_3$ the phase fractions of metallic and Mott-insulating phases was estimated using the dielectric response at finite frequency \cite{PhaseFractionPRB2021}. A similar approach could also prove useful for $\kappa$-(BEDT-TTF)$_2$Cu[N(CN)$_2$]Cl and $\kappa$-(BEDT-TTF)$_2$Hg(SCN)$_2$Br.

The exchange coupling between the electron spins in $\kappa$-(BEDT-TTF)$_2$Hg(SCN)$_2$Br can also be affected by a uniaxial strain \cite{Drichko2025}, providing another driving parameter besides the pressure. The uniaxial strain was previously used to shift the MIT in $\kappa$-(BEDT-TTF)$_2$Cu$_2$(CN)$_3$ \cite{Kawasugi2023}.

The proposed GMR mechanism is rather general and may apply to other compounds with Mott metal insulator transition, particularly with spin frustrations. For example, a similar GMR effects have been observed in another organic metal $\kappa$-(BEDT-TTF)$_2$Cu[N(CN)$_2$]Cl \cite{Kagawa2004,Oberbauer2023,Erkenov2024}, and the standard theory \cite{Laloux1994,Rozenberg1994,GeorgesRMP1996}, which neglects the glassy state, may be insufficient to explain all GMR features in this compound. Somewhat similar but a more complex pressure-temperature phase diagram appears in $\kappa$-(BEDT-TTF)$_2$Cu$_2$(CN)$_3$ \cite{Kawasugi2023,PhaseFractionPRB2021} and EtMe$_3$P[Pd(dmit)$_2$]$_2$ \cite{Shimizu2007}, where the Mott MIT also couples with spin frustrations on a triangular lattice. However, in these two organic conductors a negative magnetoresistance is observed in some parametric range \cite{Kawasugi2023,Shimizu2007}, which probably indicates a competition of two opposite effects of magnetic field on electron transport. Possibly, thermodynamically the magnetic field favors the metallic rather than insulating phase in these compounds, which leads to negative magnetoresistance, while the other effects of magnetic field increase resistivity. Note that the low-temperature Mott-insulating phase in these two compounds is suggested to have an additional charge ordering, called valence bond solid \cite{Kawasugi2023,Shimizu2007}. How this charge ordering affects magnetoresistance remains an open problem beyond the scope of our paper. In $\kappa$-(BEDT-TTF)$_2$Hg(SCN)$_2$Br and $\kappa$-(BEDT-TTF)$_2$Cu[N(CN)$_2$]Cl at low temperatures one instead observes a glassy dynamics \cite{Hemmida2018,CrystReview2018}, which couples spin and charge degrees of freedom.

 \section{Conclusion}

We observe and investigate the giant hysteretic magnetoresistance which accompanies the Mott metal-insulator transition in the organic metal $\kappa$-(BEDT-TTF)$_2$Hg(SCN)$_2$Br. The magnetoresistance and hysteresis are the strongest at low temperature and the pressure around 3$\,$kbar, corresponding to the Mott-Habbard phase transition. At our lowest temperature $T=0.5$\,K the hysteresis width reaches the interval $\Delta B\approx 4$\,T, and the ratio $R(B)/R(0)\approx 30$ in a field $B\approx 16$T. Notably, this hysteretic magnetoresistance is almost isotropic with respect to the direction of magnetic field. In particular, it is observed also for the field parallel to the conducting layers, which excludes the orbital effect of magnetic field as an origin of this phenomenon and points to its coupling to electron spin. In the organic metal $\kappa$-(BEDT-TTF)$_2$Hg(SCN)$_2$Br the behavior of spin system is complicated by antiferromagnetic frustrations on the triangular lattice, which differs our system from the standard theory \cite{Laloux1994,Rozenberg1994,GeorgesRMP1996} of Mott transition in a magnetic field. The observed temperature dependence of hysteresis, together with the relaxation time values and their dependence on magnetic field (see Fig. \ref{Fig7}) confirm the influence of spin-glass on magnetoresistance. We propose an explanation and a toy model giving qualitatively this unusual magnetoresistance behavior, which differ from the known giant magnetoresistance mechanisms. We suggest the observed effect as a novel class of extreme magnetoresistance phenomena, which couples the physics of strongly correlated electrons and GMR.

\acknowledgements{The work was partially supported by the state assignment of the FRC PCP MC RAS (registration \#124013100858-3) and by the state assignment of ISSP RAS.
P.D.G. acknowledges the state assignment \#FFWR-2024-0015.}

\section*{Data Availability}

All data are available upon reasonable request from the corresponding author at grigorev@itp.ac.ru.

\bibliographystyle{apsrev4-2}
\bibliography{Papers}


\end{document}


\title{Supplementary materials to ''Giant hysteretic magnetoresistance accompanying the Mott transition in organic metal''}

\author{P.\,D.~Grigoriev}
\affiliation{L.\,D.~Landau Institute for Theoretical Physics, 142432, Chernogolovka, Russia}
\affiliation{
	National University of Science and Technology ''MISIS'', 119049, Moscow, Russia} 
\author{S.\,I. Pesotskii}
\affiliation{FRC of Problems of Chemical Physics and Medicinal Chemistry RAS, Chernogolovka, Moscow region, 142432, Russia}
\author{R.\,B. Lyubovskii}
\affiliation{FRC of Problems of Chemical Physics and Medicinal Chemistry RAS, Chernogolovka, Moscow region, 142432, Russia}
\author{S.\,A. Torunova}
\affiliation{FRC of Problems of Chemical Physics and Medicinal Chemistry RAS, Chernogolovka, Moscow region, 142432, Russia}
\author{D.\,S.~Lyubshin}
\affiliation{L.\,D.~Landau Institute for Theoretical Physics, 142432, Chernogolovka, Russia}
\author{V.\,N.~Zverev}
\affiliation{Institute of Solid State Physics RAS, Chernogolovka, Moscow region, 142432, Russia}

\maketitle

\section{On conductivity of antiferromagnet-driven resistor lattices}

\subsection{Model description}

The dimerized BEDT-TTF molecules hosts one electron per dimer, which form a triangular lattice. The electron hopping between dimers is complicated (i) by Coulomb repulsion, when two electrons are on the same dimer, and (ii) by small interdimer transfer integral. Hence, mostly each dimer has one electron. To describe their spin configuration, we consider the Ising model on a triangular lattice. To impose periodic boundary conditions, we are working with a $L\times L$ axis-aligned parallelogram on the triangular lattice. Equivalently, we may take the square spin lattice with $N=L\times L$ sites with periodic boundary conditions, where for a site at $(x, y)$ its neighbors are the four square-lattice neighbors as well as $(x+1, y+1)$ and $(x-1, y-1)$. One classical $\pm 1$ spin is associated with each site, and the Hamiltonian reads
\begin{equation}
H=-\sum_{(ij)} J_{ij }s_i s_j - h\sum_i s_i, \qquad s_i=\pm 1, \qquad i=1,2,\dots,N,
\label{H}
\end{equation}%
where the sum is over the lattice edges. We are mostly interested in the case of purely antiferromagnetic exchange, $J_{ij}<0$. 

In the static setup, we let the system reach thermal equilibrium at temperature $T$, and then construct a resistor network with one resistor per edge; its conductivity $\sigma_{ij}$ is taken to be
\begin{equation}
\sigma_{ij} = \begin{cases} \sigma_+, & s_i s_j = +1,\\
\sigma_-, & s_i s_j = -1,
\end{cases}
	\label{sigma}
\end{equation}%
for some fixed $\sigma_+$, $\sigma_-$ (we assume $\sigma_- > \sigma_+$). The conductivity $\sigma$ of the system is then computed as the inverse resistance between the $y=0$ and $y=L$ sides of the lattice treated as line contacts. As such, the conductivity is a random variable, and for a homogeneous system ($J_{ij}=J$) its average $\sigma$ over the thermal states of the spin system depends only on two dimensionless quantities, $J/T$ and $h/T$. Most of the subsequent plots assume the ratio $J/T$ (or its analog) fixed while $h/T$ varies.

\subsection{General properties of the model}

\begin{figure}[hhtb]
	\begin{center}
		\includegraphics[width=0.78\textwidth]{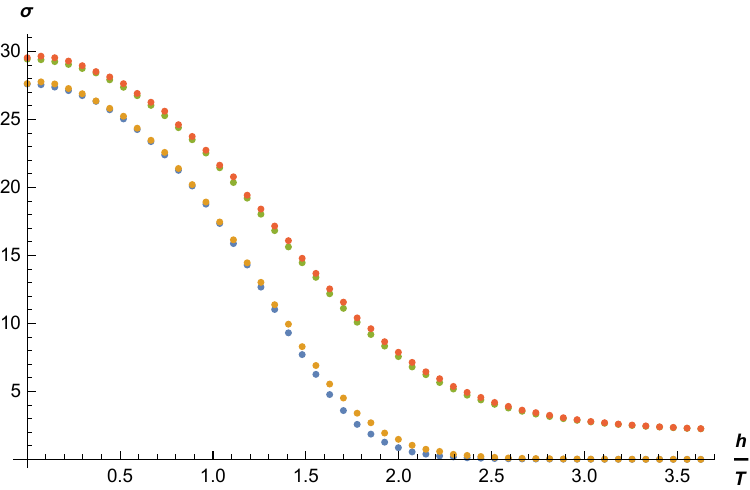}
	\end{center}	
	\caption{Calculated conductivity as a function of normalized magnetic field $h/T$, i.e. the Zeeman splitting $h=g\mu_B B$ divided by temperature $t=k_BT$ (the Boltzmann constant $k_B$ is set to unity). Here the ratio $J/T=-0.25$ is fixed. The top two curves (red and green) correspond to $\sigma_-=30$, $\sigma_+=1$, with red dots for a $L=128$ system, and green for a $L=256$ one. The bottom two curves (yellow and blue) correspond to $\sigma_-=30$, $\sigma_+=0$, with yellow obtained for a $L=128$ system, and blue for $L=256$. Each dot is a result of averaging over $10^2$ or $10^3$ samples (more for smaller system sizes).}
	\label{Fig1}
\end{figure}

A typical plot of $\sigma$ as a function of $h/T$ looks as in Fig. \ref{Fig1}.
For a homogeneous resistor network with each edge carrying conductivity $\sigma_0$ the measured value of $\sigma$ would be $2\sigma_0$ by the geometry of the triangular lattice. Hence the theoretical maximum for an observable ordinate is $2\sigma_-=60$, and in reality we have values $\sigma\approx 30$ attained at low $h/T$. As the magnetic field $h/T$ increases, more pairs of neighboring spins align and decrease the edge conductivity to $\sigma_+<\sigma_-$. For $\sigma_+=0$ the fraction of misaligned pairs eventually becomes high enough for a percolation transition to occur, setting $\sigma=0$. For $\sigma_+=1$, we get smooth asymptotic behavior instead. Naively one could expect that $\sigma$ monotonically decreases with $h/T$, but the graph shows a small initial interval of slightly increasing conductivity at low $h$ values. This is not a numerical artifact -- at lower temperatures this effect is clearly visible, with $\sigma$ remaining almost constant until the immediate vicinity of the percolation transition. 

\begin{figure}[hhtb]
	\begin{center}
		\includegraphics[width=0.78\textwidth]{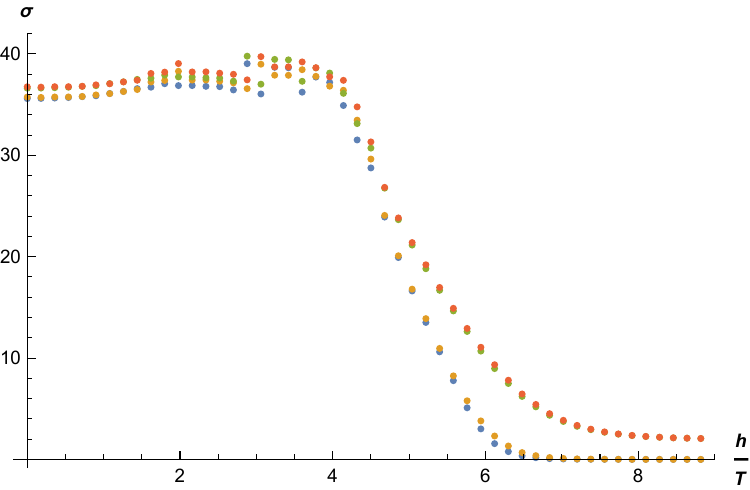}
	\end{center}
	\caption{
	The same as in Fig. \ref{Fig1} but at lower temperature, at $J/T=-1.0$.}
	\label{Fig2}
\end{figure}

It may be instructive to examine a similar plot for temperature four times lower than that in Fig. \ref{Fig1}. The corresponding result is shown in  in Fig. \ref{Fig2}. The conductivity slightly increases with $h$, then stays essentially constant over a wide range of $h$ values, entering a region of notable fluctuations for $2\leq h/T\leq 4$. This is followed with a rapid decrease of conductivity due to a percolation transition, however the fluctuations in this near-critical region unexpectedly become smaller.

\begin{figure}[hhtb]
	\begin{center}
		\includegraphics[width=0.45\textwidth]{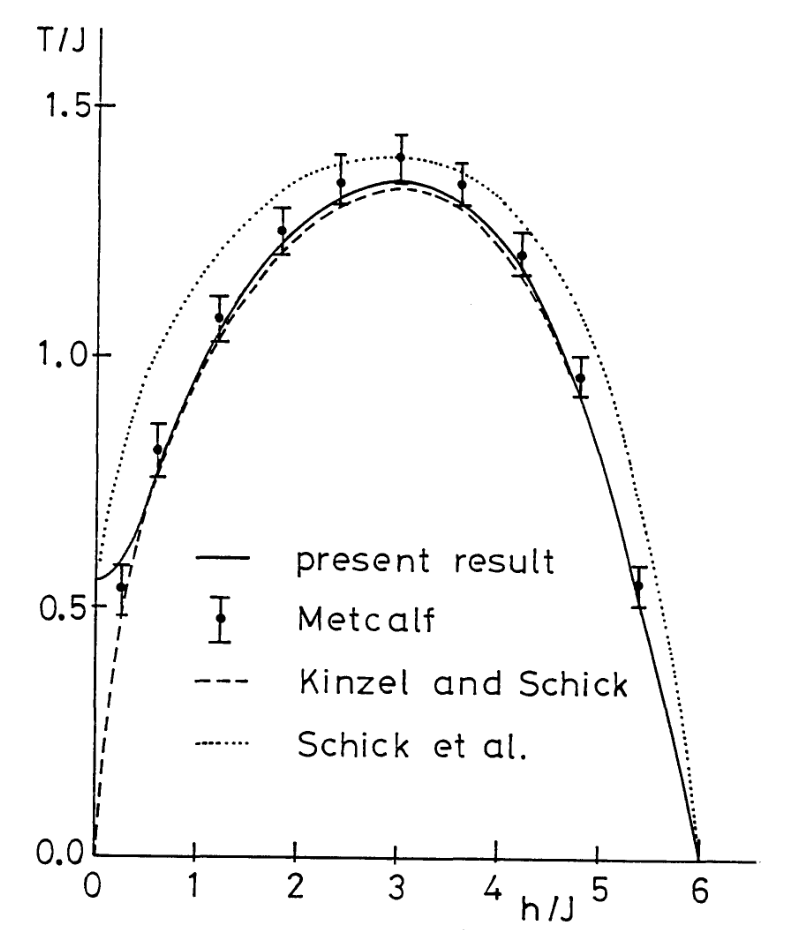}
	\end{center}	
	\caption{The phase diagram for homogeneous antiferromagnetic Ising model on a triangular lattice, taken from Fig. 6b of Ref.  \cite{SaitoIgeta1984}. A disordered (paramagnetic) phase at high temperatures or magnetic fields, and a so-called up-up-down, or AAB phase that occupies a bounded region of the phase diagram ($J/T>0.75$, $h/J<6$).}
	\label{Fig3}
\end{figure}

\begin{figure}[hhtb]
	\begin{center}
		\includegraphics[width=0.8\textwidth]{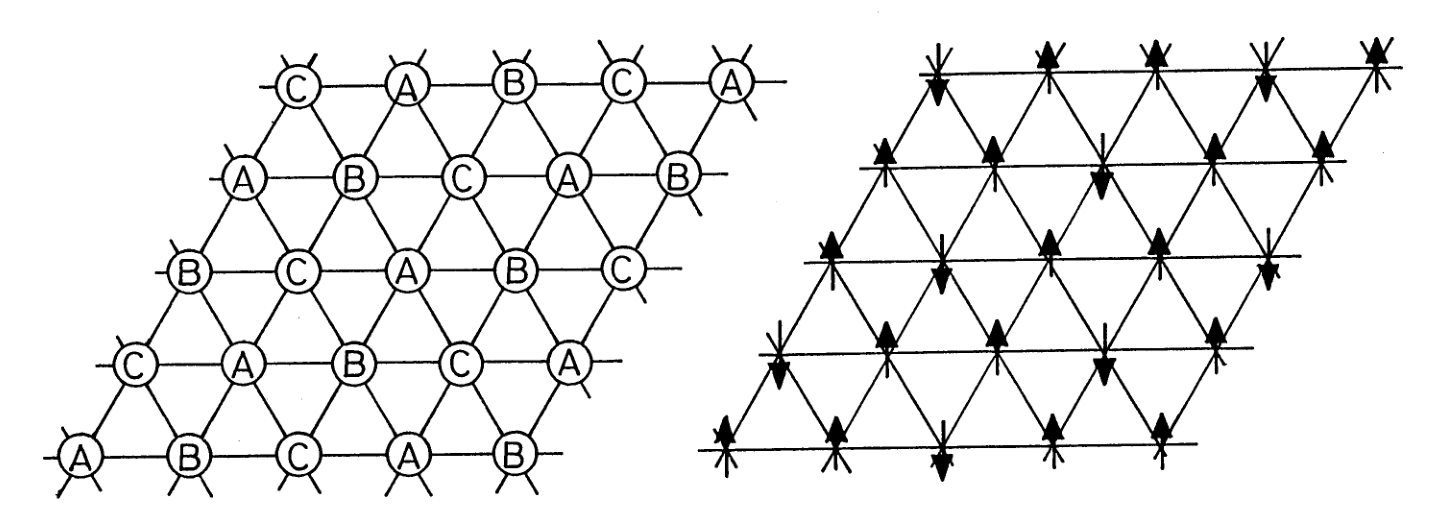}
	\end{center}
	\caption{The illustration of AAB phase of homogeneous antiferromagnetic Ising model on a triangular lattice, taken from Fig. 6b of Ref. \cite{SaitoIgeta1984}. Two of the three translationally-equivalent sublattices $A, B, C$ have equal preferred spin orientations while the third sublattice has the opposite one.}
	\label{Fig4}
\end{figure}

All these features may be easily understood theoretically if we recall the phase diagram for homogeneous model ($J_{ij}=J$) in question. The diagram features two phases (see Fig. 6 of Ref. \cite{SaitoIgeta1984}): a disordered (paramagnetic) phase at high temperatures or magnetic fields, and a so-called up-up-down, or AAB phase that occupies a bounded region of the phase diagram ($J/T>0.75$, $h/J<6$).  In the latter, two of the three translationally-equivalent sublattices $A, B, C$ have equal preferred spin orientations while the third sublattice has opposite. In Fig. \ref{Fig4} the sublattices A and B carry the up-spins, while sublattice C carries the down-spin. 

The ratio $J/T=-0.25$ of Fig. \ref{Fig1} corresponds to a horizontal line $T/J=4$ in Fig. \ref{Fig3} which stays far above the AAB phase region. For $J/T=-1$ as in Fig. \ref{Fig2}, the system enters the AAB phase at about $h/T=1.25$. It then traverses the $T/|J|=1$ line inside this phase, all the time staying relatively close to the top part of phase boundary (this line of second order transitions is responsible for the stronger fluctuations), and exits the AAB phase at $h/|J|\approx 4.5$. While in the AAB phase, the site averages favor an almost fixed resistor configuration illustrated in Fig. \ref{Fig4}. The percolation transition happens while the system is already back in the low-fluctuating paramagnetic phase.  

This may be confirmed by plotting the average magnetization $\langle s_i\rangle$ as a function of $h/T$ (see Fig. 5). The solid horizontal line corresponds to the theoretical value $\langle s_i\rangle=(1+1-1)/3=1/3$ in the saturated AAB phase. All four colors of Fig.2 are shown, but largely collapse. Note that the discrepancy between $L=128$ and $L=256$ results may be taken for an estimate of general reliability of the values, with combined statistical and systematic errors being of the order the shown discrepancy. The conductivity for a pure AAB state is $\sigma=40.0$, also is a reasonable agreement with Fig. \ref{Fig2}.

\begin{figure}[hhtb]
\begin{center}
\includegraphics[width=0.8\textwidth]{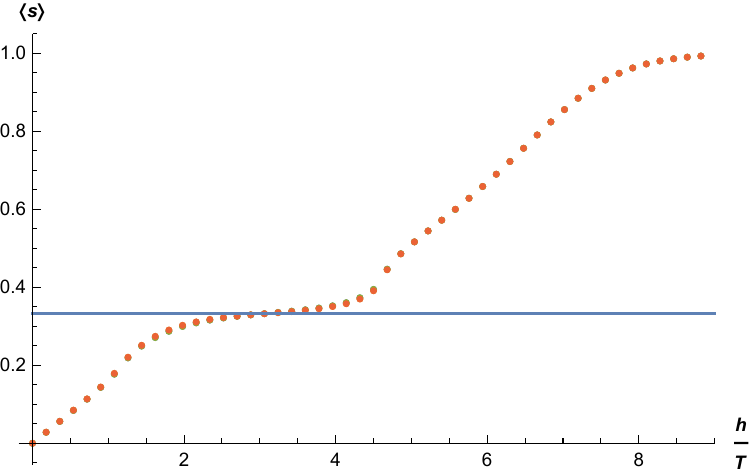}\\
\end{center}
	\caption{ {\bf  The average doubled spin projection per site $\langle s_i\rangle$, giving magnetization, as a function of $h/T$.} The solid blue horizontal line corresponds to the theoretical value $\langle s_i\rangle=(1+1-1)/3=1/3$ in the saturated AAB phase.}
	\label{Fig5}
\end{figure}

\begin{figure}[hhtb]
\begin{center}
\includegraphics[width=0.8\textwidth]{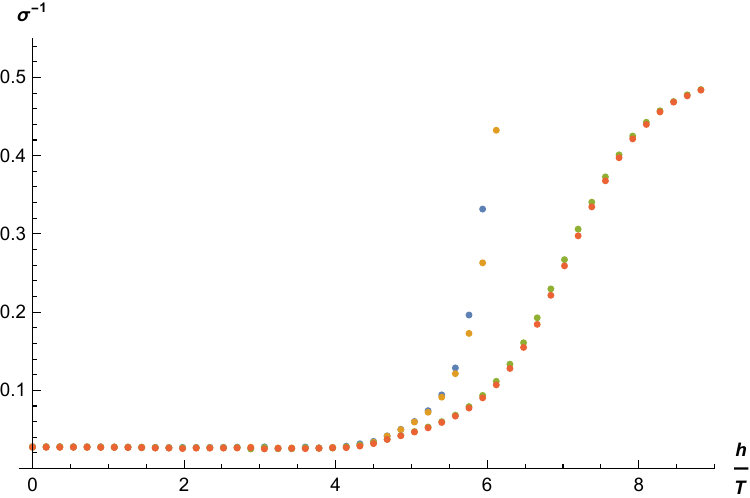}\\
\end{center}
\caption{Resistivity $\rho=\sigma^{-1}$ as a function of magnetic field at the same parameters as in Fig. \ref{Fig2}.}
\label{Fig6}
\end{figure}

It's worth noting that the plots of resistivity $\rho=\sigma^{-1}$ are much less sensitive to these finer details. For example, Fig.2 corresponds to the following resistivity plot shown in Fig. \ref{Fig6}. 

Similar plots are expected for all sufficiently low temperatures. They all show giant magnetoresistance accompanying the metal-insulator transition. At high temperature this strong magnetoresistance starts from low magnetic fields (see Fig. \ref{Fig1}). However, at low temperature in our triangle resistor lattice driven by the classical antiferromagnetic Ising model the giant magnetoresistance is more limited to the range of metal-insulator transition (see Fig. \ref{Fig6}). This is because in the highly conductive ''metallic'' region our model does not include the orbital effect of magnetic field leading to usual positive magnetoresistance. The latter appears even if the magnetic field is parallel to the electric current across the conducting layers \cite{Gvozdikov2007,Kartsovnik2009Apr,Grigoriev2011Jun,Grigoriev2012Oct,Grigoriev2013Aug}. 

\newpage

\subsection{Hysteresis without disorder in the Ising model}

Of course, in the truly static case described above there may be no hysteresis of any kind. To obtain the hysteresis, we need to specify a time-dependent process that does not allow complete thermalization. We always consider a measurement scheme in which a system, initially held at $h=0$, first undergoes a rapid increase of $h$ to some positive values above the percolation threshold, followed by thermalization at high $h$, and then a rapid decrease of $h$ back to zero values. The temperature $T$ (and, of course, the interaction parameters $J_{ij}$) are held constant. 

In such a setting, the meaning of the horizontal axis becomes twofold: it is both a magnetic field ($h/T$) axis and a time axis (swept in two opposite directions). A measure of rapidity of the field change is the amount of time spent at each fixed value of $h$. Since the partial thermalization is achieved by running a Metropolis-Hastings algorithm with randomly chosen updated sites, the time is naturally measured in the average number $N_u$ of spin update attempts at a fixed site before the $h$ value changes. The fastest meaningful process has $N_u\approx 1$, while $N_u\approx 10L$ is already sufficiently slow in the context of this model.   

\begin{figure}[hhtb]
\begin{center}
\includegraphics[width=0.8\textwidth]{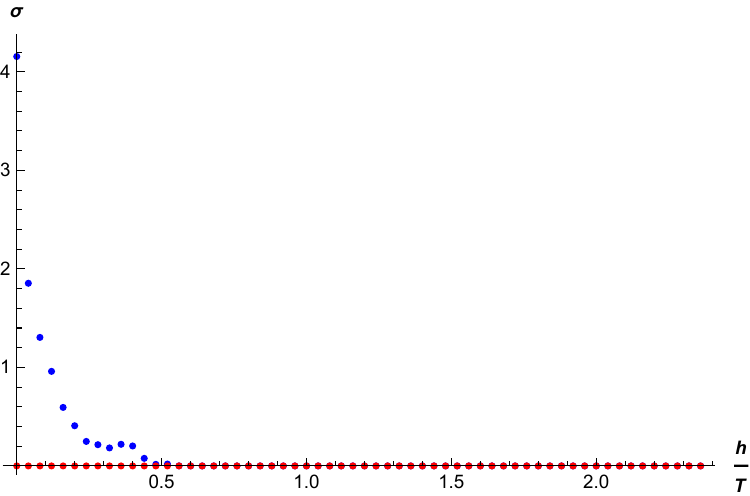}\\
{\bf Fig.7}
\end{center}
\caption{Conductivity $\sigma$ as a function of magnetic field in a {\it ferromagnetic} resistor lattice at $J/T=+0.25$ close to the ferromagnetic transition point $J_c/T=\log(3)/4 \approx 0.275$. Blue dots correspond to increasing magnetic field $h/T$ and the red dots to decreasing $h/T$.}
	\label{Fig7}
\end{figure}

As an illustration, consider a ``trivial'' run of a {\it ferromagnetic} system, illustrated in Fig. \ref{Fig7}. Here $J/T=+0.25$, which is pretty close to the ferromagnetic transition point $J_c/T=\log(3)/4 \approx 0.275$. The magnetic field change is maximally fast, with on average only $N_u=1$ spin update at fixed $h$. Blue dots correspond to the forward motion (increasing $h/T$), and the red dots correspond to the reverse (decreasing $h/T$). The system rapidly reaches a ferromagnetic state, and the horizontal axis is much better interpreted as time than as magnetic field here: in approximately 12 iterations, the ferromagnetic state with $\sigma=0$ is reached and never exited afterwards. This is very rapid mixing. 

\begin{figure}[hhtb]
\begin{center}
\includegraphics[width=0.8\textwidth]{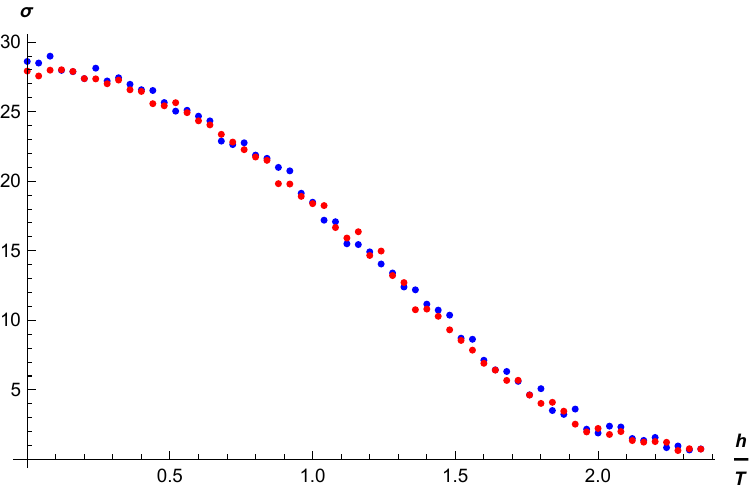}\\
{\bf Fig.8}
\end{center}
\caption{Conductivity $\sigma$ as a function of magnetic field in antiferromagnetic resistor lattice at $J/T=-0.25$ for increasing (blue dots) and decreasing (red dots) magnetic field $h/T$.}
	\label{Fig8}
\end{figure}

With these remarks in mind, we return to the antiferromagnetic system. The plot in Fig. \ref{Fig8} corresponds to the same schedule, except with $J/T=-0.25$. Only the $\sigma_+=0$ curves are shown; as before, blue dots mean increasing $h/T$, while red dots mean decreasing $h/T$. The fully thermalized counterpart of this Fig. \ref{Fig8} is Fig. \ref{Fig1}. Complete thermal equilibrium was achieved at $h/T=0$, then we proceeded with the maximally rapid field increase with $N_u=1$ spin update per $h/T$ value; then a long thermalization at the highest value of $h/T$ is followed by a similarly rapid decrease in $h$. 

No traces of hysteresis may be seen in Fig. \ref{Fig8}. Although the figure corresponds to just one sample (no averaging), any other single sample produces an essentially the same curve. This qualitative picture is also representative for all sufficiently high temperatures (far above the AAB phase region). As with the ferromagnetic case, the mixing is very rapid. 

\begin{figure}[hhtb]
\begin{center}
\includegraphics[width=0.8\textwidth]{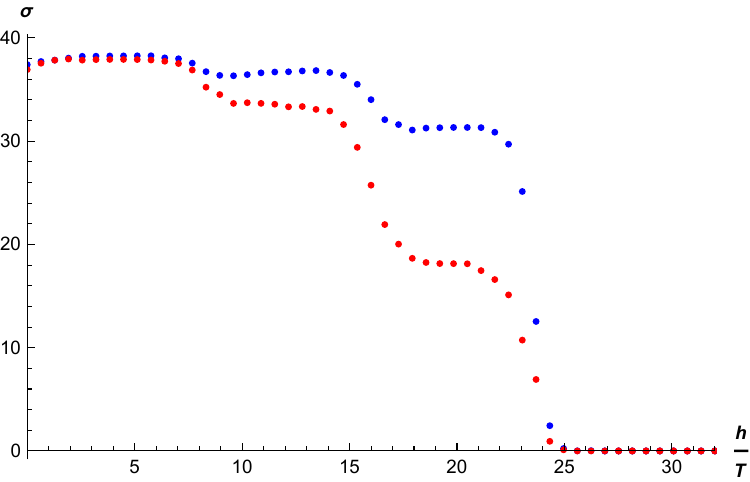}\\
{\bf Fig.9}
\end{center}
\caption{Conductivity $\sigma$ as a function of magnetic field in antiferromagnetic resistor lattice at low temperature $J/T=-4$ for increasing (blue dots) and decreasing (red dots) magnetic field $h/T$.}
\label{Fig9}
\end{figure}

For visible hysteresis, we must consider very low temperatures, as illustrated in Fig. \ref{Fig9}. Here $J/T=-4$, and otherwise the parameters are the same as for Fig. \ref{Fig8}. Understandably, the field-reverse (red) part has smaller conductivity. Although the curves look arbitrary (say, one can expect the positions of almost horizontal regions to be random, reflecting stochastic jumps to better local extrema), the picture is in fact stable with respect to reproduction in samples with different initial states, random seeds governing Metropolis spin updates, etc.  

\begin{figure}[hhtb]
	\begin{center}
		\includegraphics[width=0.8\textwidth]{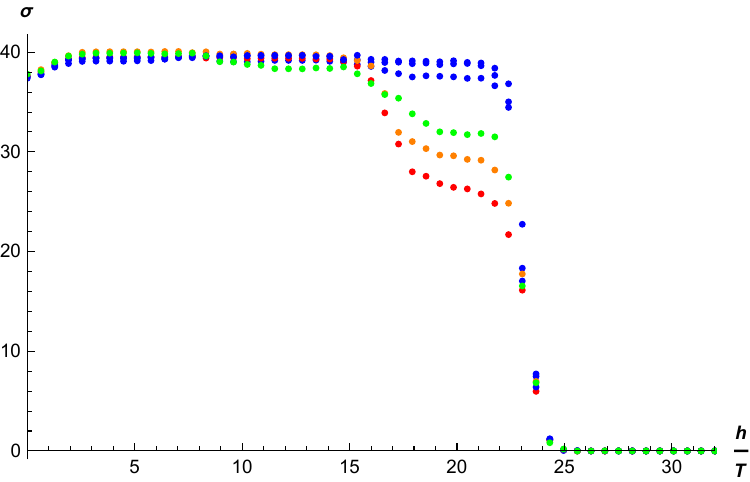}\\
		{\bf Fig.10}
	\end{center}
	\caption{Conductivity $\sigma$ as a function of magnetic field in antiferromagnetic resistor lattice at $J/T=-4$ for increasing (blue dots) and decreasing magnetic field $h/T$ at three different speeds: $N_u=50$ (red), $N_u=150$ (orange), and $N_u=256$ (green) dots.}
	\label{Fig10}
\end{figure}

Although we got a reproducible hysteresis loop, this is very much an artifact of extra-accelerated field change schedule. Unlike in a typical spin glass, the barriers involved here are not infinite in the thermodynamic limit, and thermalization times do not scale exponentially with system size. As we shift from the minimal $N_u=1$ spin update for fixed field to higher values, the gap between the blue and red curves closes, as illustrated in Fig. \ref{Fig10}.

The plots in Fig. \ref{Fig10} correspond to the same value $J/T=-4$, but shows three different speeds of varying the field: $N_u=50$ (red), $N_u=150$ (orange), and $N_u=256$ (green) dots for the reverse part. The forward part (increasing $h/T$) is shown as blue for all three schedules, creating a clustered top curve. As the thermalization time increases, the hysteresis loop closes. Note that again we get a perfect agreement with the prediction of constant conductivity throughout the AAB phase: not only the ordinate is indeed $\sigma=40$, but also the termination point is $(h_{\rm max}/T)/(|J|/T)=6$ matching the low-temperature bound $h/J=6$ of Fig. \ref{Fig3}.

The above data suggest that the resistor lattice driven by the classical antiferromagnetic Ising model subject to an external field produces results 
different from the experimental plots of magneto\-resistance. This model indeed gives both the giant magnetoresistance accompanying the metal-insulator transition and its hysteresis, as observed in our experiment. However, the size of this hysteresis is smaller than the experimentally observed and depends on the speed of field change. In addition, the range of giant magnetoresistance in the resistor lattice driven by the classical antiferromagnetic Ising model is rather close to the metal-insulator transition, while in experiment the giant magnetoresistance has a wider range.  

There are two regimes of the classical antiferromagnetic Ising model: (1) the high temperature case when the system remains paramagnetic, and (2) the low-temperature regime where the system mostly stays in the AAB phase, with an early entry into it at low $h$, and a late exit shortly before the percolation transition. 
In phase (1) the mixing is extremely fast, with no chance for any observable hysteresis. In phase (2) observing a weak hysteresis confined to a narrow region of the magnetic fields is in principle possible, but the plot of the conductivity has a very distinctive form with a large plateau followed with a very narrow region of rapid change and subsequent complete saturation. Neither picture matches quantitatively the experimental plots. 

\subsection{Hysteresis in the resistor lattice driven by antiferromagnetic Ising model with disorder}

The situation becomes somewhat better if we introduce significant quenched disorder, making the inter\-action constants $J_{ij}$ random. For example, consider the model with $J_{ij}$ being i.i.d.~random variables uniform in $[-2|J|, 0]$ (i.e.~of box distribution with half-width $|J|$ centered at $-|J|$). We can repeat the dynamic schedule of Fig.8 to obtain Figs. 11 and 12.

\begin{figure}[hhtb]
\begin{center}
\includegraphics[width=0.8\textwidth]{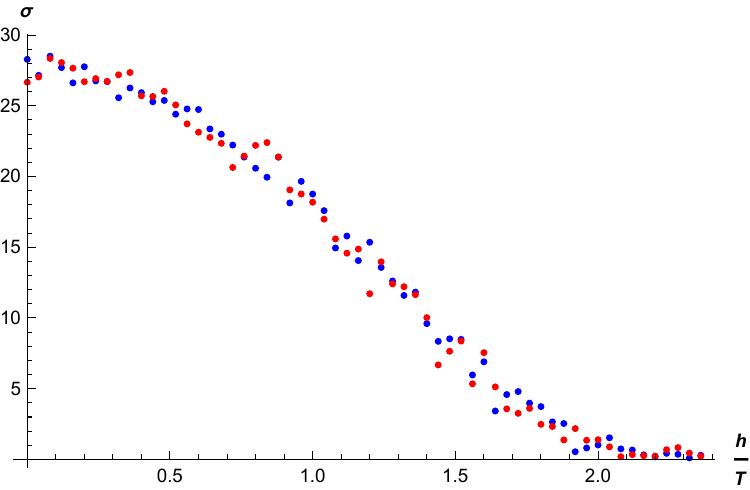}
\end{center}
	\caption{Conductivity $\sigma$ as a function of increasing (blue dots) and decreasing (red dots) magnetic field $h/T$ in antiferromagnetic resistor lattice at high temperature $J/T=-0.25$  and random interaction constants $J_{ij}\in [-2|J|, 0]$.}
	\label{Fig11}
\end{figure}

\begin{figure}[hhtb]
	\begin{center}
		\includegraphics[width=0.8\textwidth]{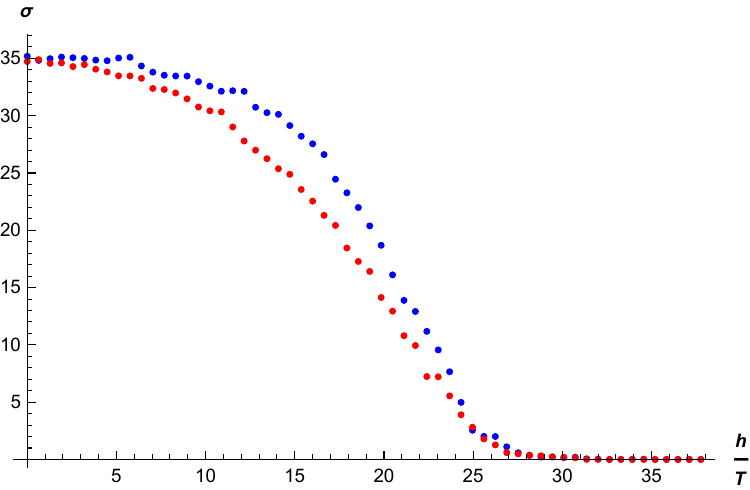}
	\end{center}
	\caption{Conductivity $\sigma$ as a function of increasing (blue dots) and decreasing (red dots) magnetic field $h/T$ in antiferromagnetic resistor lattice at low temperature $J/T=-4$ and random interaction constants  $J_{ij}\in [-2|J|, 0]$.}
	\label{Fig12}
\end{figure}

As we see from Fig. \ref{Fig11}, at high temperatures no major changes occur: even a very fast schedule of varying the magnetic field leads to rapid mixing and absence of any hysteresis. But for low temperatures the form of the plot is modified; repeating the dynamic schedule of Fig. \ref{Fig9} (and keeping all the respective parameters intact, except for now inhomogeneous $J_{ij}$ values), we get the results shown in Fig. \ref{Fig12}.

The hysteresis curve is now smoother and sweeps the whole range of magnetic fields; also, the form of the conductivity curve is more regular and does not exhibit the highly specific plateau. While the extended hysteresis range may be a manifestation of a more complex structure of local minima of spin glass type, the change of the general shape is likely due simply to the destruction of the AAB phase region, making the system more generic. 

\begin{figure}[hhtb]
\begin{center}
\includegraphics[width=0.80\textwidth]{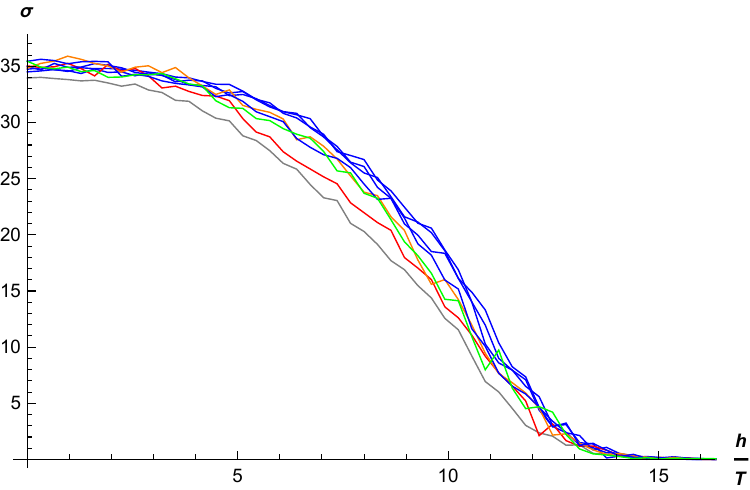}
\end{center}
	\caption{Conductivity $\sigma$ as a function of increasing (blue lines) and decreasing (other colors) magnetic field $h/T$ in antiferromagnetic resistor lattice at moderately low temperature and random interaction constants $J/T=-2.0\pm 2.0$. Four different schedules (speeds of magnetic field sweep) differ by the average number of attempted flips per spin: $1$ (gray), $10$ (red), $10^2$ (green), and $10^3$ (orange).}
\label{Fig13}
\end{figure}

We now investigate the properties of the hysteresis loop in the disordered system, addressing the dependence on the rapidity of the field change schedule and the strength of the disorder. A typical picture for strong disorder and moderately low temperatures is as shown in Fig. \ref{Fig13}. This plot corresponds to $J/T=-2.0\pm 2.0$ (uniform distribution) and field sweeps $h/T = 0 \to 20.0 \to 0$. Four different schedules differ by the average number of attempted flips per spin ($1$, $10$, $10^2$, $10^3$). A characteristic feature of strong disorder is the very weak dependence of conductivity on the schedule in the first (forward, increasing $h/T$) half of the sweep; to avoid extra cluttering of the plot, all four variants have forward parts of the plots given in blue. The backward half of the sweep displays a hysteresis loop with magnitude decreasing for longer wait times: the plots for the four schedules are shown as gray, red, green, and orange (in this order as the wait times increases). For the larger of those, we essentially see the loop closed.

\begin{figure}[hhtb]
\begin{center}
\includegraphics[width=0.80\textwidth]{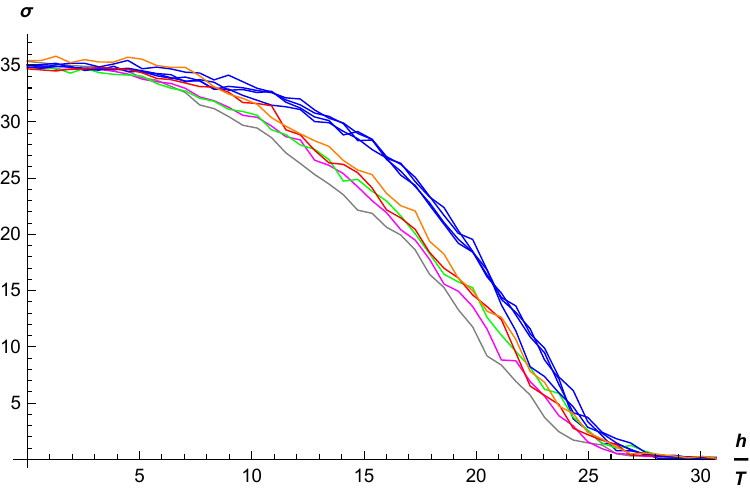}
\end{center}
	\caption{Conductivity $\sigma$ as a function of increasing (blue dots) and decreasing (red dots) magnetic field $h/T$ in antiferromagnetic resistor lattice at low temperature $J/T=-4$ and random interaction constants $J_{ij}$.}
\label{Fig14}
\end{figure}

A typical picture for strong disorder and low temperatures is given in Fig. 14. Here $J/T=-4.0\pm 4.0$, so compared to Fig.13 we have the same system at twice lower temperature. Five field change schedules correspond to 1 (gray), 10 (magenta), $10^2$ (green), $10^3$ (red), $10^4$ (orange) attempted flips per spin; a scale of four orders of magnitude is insufficient to fully close the hysteresis loop. At this temperature it remains visible for all  waiting times we tried (and will likely stay visible for all realistically achievable times), matching the expected glass-like behavior. Recall that the system without disorder lost all signs of the hysteresis for the largest waiting times used here.

\begin{figure}[hhtb]
	\begin{center}
		\includegraphics[width=0.78\textwidth]{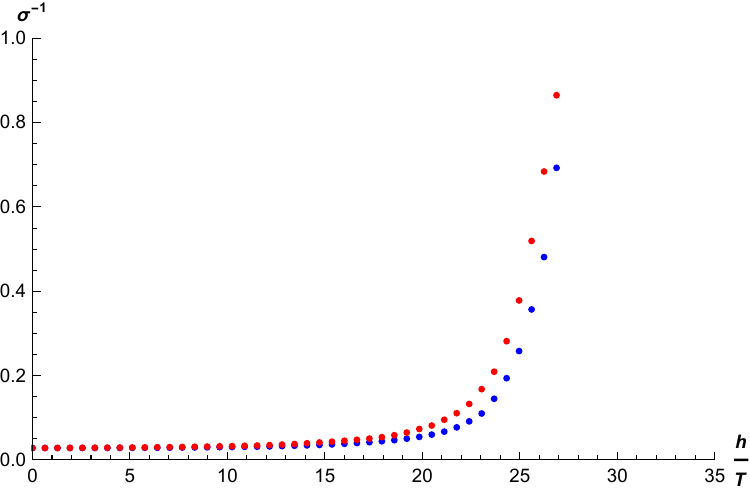}\\
		{\bf Fig.15}
	\end{center}
	\caption{Resistivity $\rho =\sigma^{-1}$ as a function of increasing (blue dots) and decreasing (red dots) magnetic field $h/T$ in antiferromagnetic resistor lattice at low temperature $J/T=-4$ and random interaction constants $J_{ij}\in [-2|J|, 0]$.}
	\label{Fig15}
\end{figure}

It should be mentioned, however, that the hysteresis is much less prominent on the plots for the resistivity. Plotting the data corresponding to the parameter set of Fig. \ref{Fig12} with reciprocals along the vertical axis results in figures like shown in Fig. \ref{Fig15}. As before, the forward sweep is shown in blue, the backward sweep in red. This qualitative behavior is quite expected: the point $h=0$ is in no way special for the model, which is deep in the supercritical state as a percolation system. Only in the narrow region near the percolation phase transition the resistivity changes significantly, and this change has a type of a singularity. It is unlikely that any tuning of the parameters of the classical resistor model can result in a strong power-law type dependence of the resistivity of the magnetic field in the whole range of field values. While this power-law behavior may be attributed to other mechanisms \cite{Gvozdikov2007,Kartsovnik2009Apr,Grigoriev2011Jun,Grigoriev2012Oct,Grigoriev2013Aug} of magnetoresistance that are additive with the one considered, we would then be forced to explain the observed wide hysteresis loop that persists at all values of the field by other mechanisms as well, making quantitative relevance of the coupled spin glass and resistor network mechanism questionable. For a quantitative comparison with experimental data one should go beyond the considered classical resistor lattice driven by antiferromagnetic Ising model. Probably, a calculation using the dynamic mean-field theory (DMFT) or other techniques developed for the analysis of many-particle problems are more suitable for $\kappa$-(BEDT-TTF)$_2$Hg(SCN)$_2$Br, but more complicated and less intuitive.

\begin{figure}[hhtb]
\begin{center}
\includegraphics[width=0.77\textwidth]{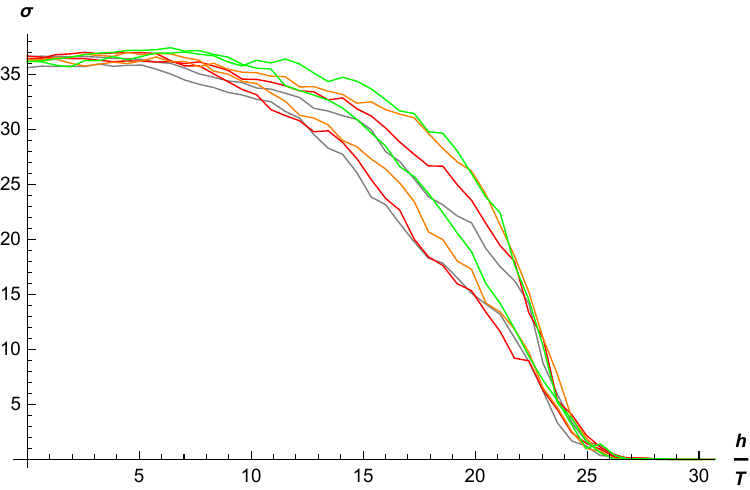}\\
{\bf Fig.16}
\end{center}
	\caption{Conductivity $\sigma$ as a function of increasing and decreasing magnetic field $h/T$ in antiferromagnetic resistor lattice at low temperature $J/T=-4$ and random interaction constants $J_{ij}/T=-4.0\pm 2.0$. The four schedules (with 1, 10, $10^2$, $10^3$ attempted flips per spin), corresponding to different rates of magnetic field change, are now shown respectively in gray, red, orange, and green for both directions of the field sweep.}
\label{Fig16}
\end{figure}

\begin{figure}[hhtb]
\begin{center}
\includegraphics[width=0.77\textwidth]{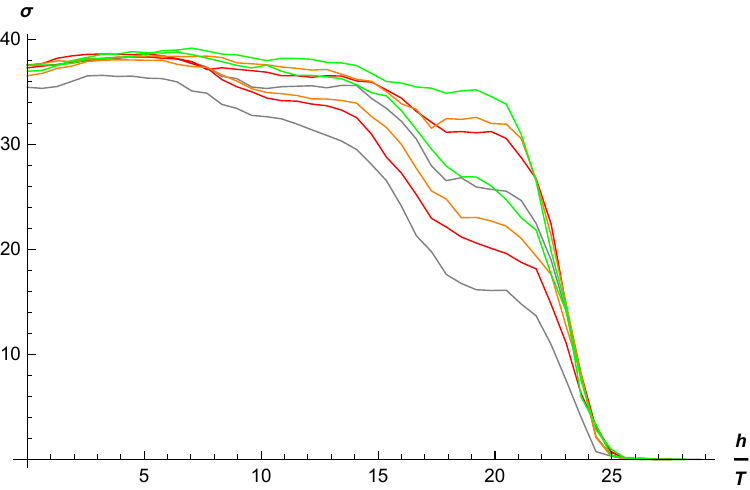}\\
{\bf Fig.17}
\end{center}
	\caption{The same as in Fig. \ref{Fig16} but at smaller disorder of interaction constants $J_{ij}/T=-4.0\pm 1.0$.}
\label{Fig17}
\end{figure}

Finally, we demonstrate the interpolation between ordered and strongly disordered systems. The plots in Fig. \ref{Fig16} and Fig.  \ref{Fig17} correspond to $J/T=-4.0\pm 2.0$ and $J/T=-4.0\pm 1.0$ respectively, i.e.~to the same temperature and average interaction constant as that in Fig. \ref{Fig14}, but with the variation in the interaction constants twice or four times smaller. 
A notable difference is a much stronger dependence of the conductivity during the forward sweep on the field change schedule. Thus the four schedules (with 1, 10, $10^2$, $10^3$ attempted flips per spin) are now shown respectively in gray, red, orange, and green for both directions of the sweep. The forward sweep corresponds to the upper curve (higher values of the conductivity). The qualitative changes in the strength of the hysteresis loops are not entirely obvious from the plot, but the situation clarifies if we reduce the strength of the disorder further.

In Fig. \ref{Fig17} we have $J/T=-4.0 \pm 1.0$, while the meaning of colors is the same as in Fig. \ref{Fig16}. We see that the forward curve tends to the (almost) horizontal line followed by a rapid drop, while the range of the fields where the hysteresis loop is wide becomes further limited to the vicinity of the percolation phase transition. Comparison with Fig. \ref{Fig10} shows that we are now in the weak disorder regime, completing the ordered-disordered interpolation. 

\bibliographystyle{apsrev4-2}
\bibliography{Papers}
